\documentclass[aip,preprint]{revtex4-1}
\usepackage{svg}
\usepackage{caption}
\usepackage{subcaption}
\usepackage{siunitx}

\usepackage{soul}
\usepackage{xcolor}
\usepackage{comment}
\draft 

\begin{document}


\title{Spanwise Control Authority of Synthetic Jets on a Stalled Airfoil}

\author{Adnan Machado}
\thanks{Author to whom correspondence should be addressed: adnan.machado@mail.utoronto.ca}
\affiliation{Department of Mechanical and Industrial Engineering, University of Toronto, Toronto, Ontario, M5S 3G8, Canada}

\author{Kecheng Xu}
\affiliation{University of Toronto Institute for Aerospace Studies, Toronto, Ontario, M3H 5T6, Canada}

\author{Pierre E. Sullivan}
\affiliation{Department of Mechanical and Industrial Engineering, University of Toronto, Toronto, Ontario, M5S 3G8, Canada}

\date{\today}

\begin{abstract}
This study investigates the aerodynamic effects of low- and high-frequency synthetic jet control strategies on a National Advisory Committee for Aeronautics (NACA) 0025 airfoil. Visualizations and measurements are employed to assess the stability of the flow, focusing on the shear layer and wake dynamics under two forcing frequencies. High-frequency actuation is found to induce steadier flow reattachment and more favorable aerodynamic characteristics compared to low-frequency control. Flow structures resulting from high-frequency actuation, notably vortex rings, are identified and their significance in flow control is evaluated. The spanwise control authority of the synthetic jet array is evaluated, revealing that the aerodynamic stability decreases significantly away from the midspan. Additionally, the effective control length is limited to approximately 40\% of the length of the array. Insights from modal analysis provide additional understanding of flow structures and their evolution across different spanwise planes.
\end{abstract}


\maketitle 

\section{Introduction}
\label{sec:intro}
Airfoils with short chord lengths are prone to stall due to flow separation, particularly at low speeds and high altitudes, resulting in a significant loss of lift and increased drag. Flow separation critically impacts aerodynamic efficiency and imposes substantial constraints on the operational envelope of airfoils at low Reynolds numbers ($\mathrm{Re}_c<10^6$). These challenges are evident across various applications, from drones designed for surveying and stealth operations, to electric planes~\cite{Lefebvre2015}, and extending to wind turbines~\cite{Melius2016}.

Flow control methods can be employed to manipulate the flow over a wing to delay separation and prevent stall. While it is possible to use either active or passive flow control, passive control requires physical changes to the aerodynamic surface causing parasitic drag. Synthetic jet actuators (SJAs) are zero-net-mass-flux devices that add momentum to the flow via periodic suction and blowing~\cite{Smith1998}. As these devices become more compact, lightweight, and energy-efficient, SJAs have promising applications to flow control. SJAs can reattach separated flows by enhancing mixing between the freestream and the separated shear layer. This mixing results in downward momentum transfer, energizing the shear layer and ultimately reattaching the flow.

The actuation frequency of SJAs is a crucial control parameter for effective flow control, as it dictates the formation of flow structures that serve as the physical mechanisms driving mixing and momentum transfer. However, SJAs must operate near their resonant frequency to maximize momentum flux. Burst modulation, achieved by intermittently activating an SJA at its resonant frequency, enables the targeting of global instabilities associated with significantly lower frequencies. Burst modulation has proven to be an effective technique in flow control, as demonstrated by various experimental studies~\cite{Glezer2005,Feero2015,Feero2017a,Feero2017b,Rice2018,Rice2021,Yang2022,Kim2022,Xu2023,Machado2024}. Additionally, this technique has the benefit of reduced power consumption, due to the SJA only being powered for a fraction of the signal's period. The modulation frequency, $f_m$, is nondimensionalized similarly to the Strouhal number, as $F^+=f_mc/U_\infty$, where $c$ is the chord length and $U_\infty$ is the freestream velocity. Prior experimental investigations have focused on two distinct frequency regimes: a) low-frequency actuation at $F^+\approx\mathcal{O}(1)$, and b) high-frequency actuation at $F^+\approx\mathcal{O}(10)$. Low-frequency actuation, which targets the natural shedding frequency in the wake of the airfoil ($\mathrm{St_w}\approx\mathcal{O}(1)$), has been extensively studied for its ability to induce large-scale spanwise vortices and enhance fluid entrainment from the freestream. This process energizes the separated shear layer and facilitates flow reattachment, leading to a significant increase in lift~\cite{Greenblatt2000,Amitay2002,Glezer2005,Salunkhe2016,Feero2017b,Kim2022,Yang2022,Xu2023}. However, despite its well-documented benefits for lift enhancement, studies suggest that high-frequency actuation offers superior drag reduction~\cite{Amitay2001,Amitay2002, Glezer2005,Feero2017b,Xu2023} and steadier aerodynamic forces~\cite{Wu1998, Feero2015,Machado2024} over low-frequency actuation.



The orifice geometry of the SJA strongly influences the types of flow structures generated. Rectangular slot style SJAs have been studied extensively for their application in flow control due to their advantage in fluid entrainment~\cite{Toyoda2009}. However, these SJAs require significant modification to the wing surface and often require large cavities, introducing structural concerns. Alternatively, using an array of microblowers with small circular nozzles presents a practical design solution more viable for engineering applications. MEMS-based microblowers also have the advantage of low power requirements and minimal noise production compared to large cavity SJAs~\cite{Xu2023}. While the effect of circular SJAs on simple geometries has been studied, such as flat plates and circular cylinders, only limited experiments~\cite{Tang2014, Salunkhe2016,Xu2023,Machado2024} have studied their application in flow control for airfoils.

Prior research has predominantly studied the impact of flow control at the midspan of the wing. However, it is also of interest to understand the three-dimensional effects of flow control with an array of SJAs. Recent visualizations~\cite{Machado2024} revealed flow convergence towards the center, accompanied by significant spanwise velocities. Interestingly, the effective spanwise extent of effectively controlled flow was found to be much narrower than the SJA array's length. Similar flow convergence patterns were observed using a rectangular, slot-shaped SJA in oil flow visualization experiments~\cite{Feero2017a}, as well as in an experimental and numerical study~\cite{Sahni2011} in which significant spanwise velocity components were measured. A comprehensive understanding of the three-dimensional flow dynamics is essential for improving the spanwise control authority.

This paper begins by comparing the aerodynamic effects of low- and high-frequency synthetic jet actuation on a stalled airfoil, with a particular focus on the stability of the shear layer and the wake. Next, we aim to explain the effectiveness of high-frequency actuation by identifying the induced flow structures and evaluating their significance in flow control. Lastly, we investigate the spanwise control authority by a) characterizing how the aerodynamic performance deteriorates away from the midspan, and b) analyzing the evolution of flow structures at various spanwise planes.

\section{Experimental Method}
\label{sec:expmeth}
\begin{figure}
    \centering
    \includegraphics[width=0.5\linewidth]{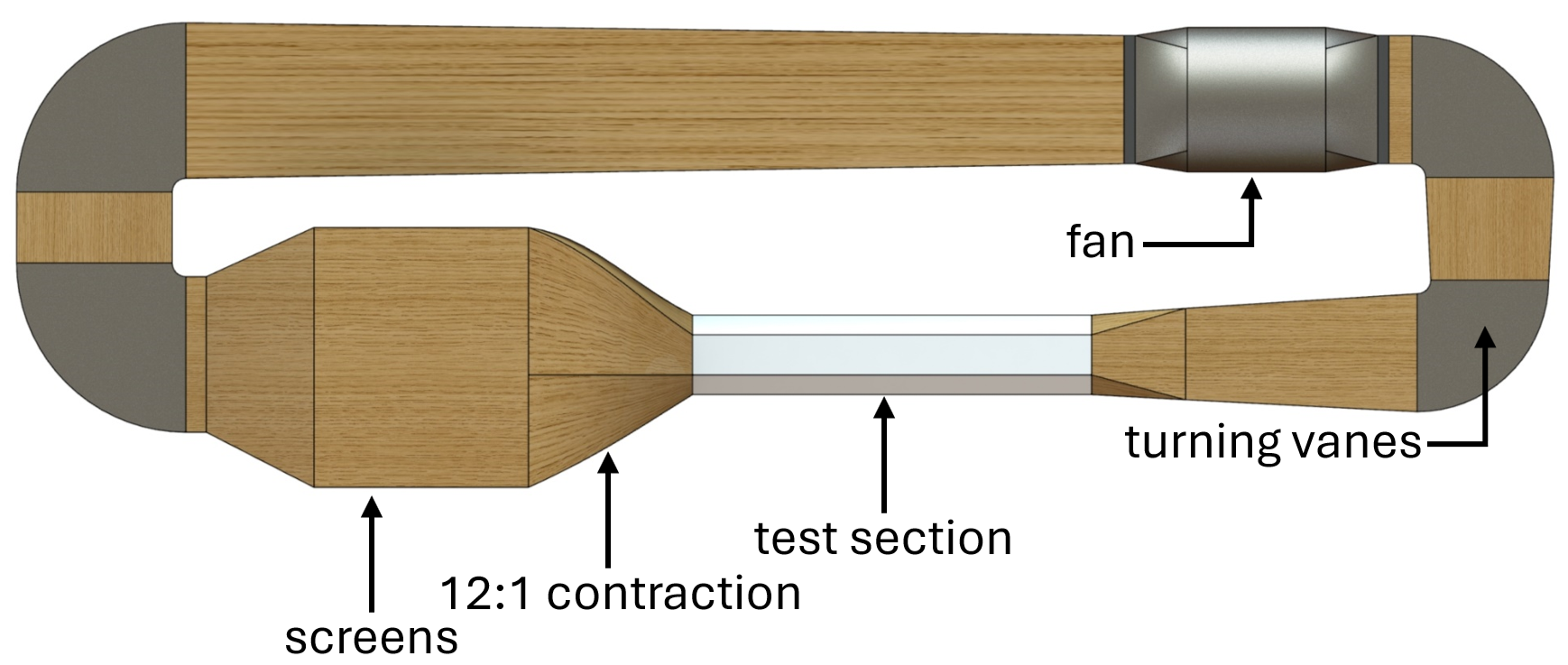}
    \caption{Labelled schematic of the wind tunnel}
    \label{fig:wind tunnel}
\end{figure}

Experiments were conducted in the low-speed recirculating wind tunnel at the Department of Mechanical and Industrial Engineering at the University of Toronto (Fig.\ref{fig:wind tunnel}). The test section has dimensions of 5~\unit{\metre}~$\times$~0.91~\unit{\metre}~$\times$~1.22~\unit{\metre} and features acrylic windows on the top and side walls for observation and measurement. The flow passes through seven screens and a 12:1 contraction before entering the test section. The wind tunnel can produce speeds between 3--18~\unit[per-mode = symbol]{\metre\per\second} with a turbulence intensity of less than 1\%. The freestream velocity was measured with a pitot-static tube at the test section entrance (35~\unit{\centi\metre} above and 28~\unit{\centi\metre} upstream of the leading edge of the airfoil) connected to a differential pressure transducer. The uncertainty of the freestream velocity is estimated to be less than $\pm1\%$. For the experiments conducted, the wind tunnel was operated at a freestream velocity of $U_\infty=5.1$~\unit[per-mode = symbol]{\metre\per\second}, resulting in a chord-based Reynolds number of $\mathrm{Re}_c=10^5$.

\begin{figure}
    \centering
    \includegraphics[width=0.5\linewidth]{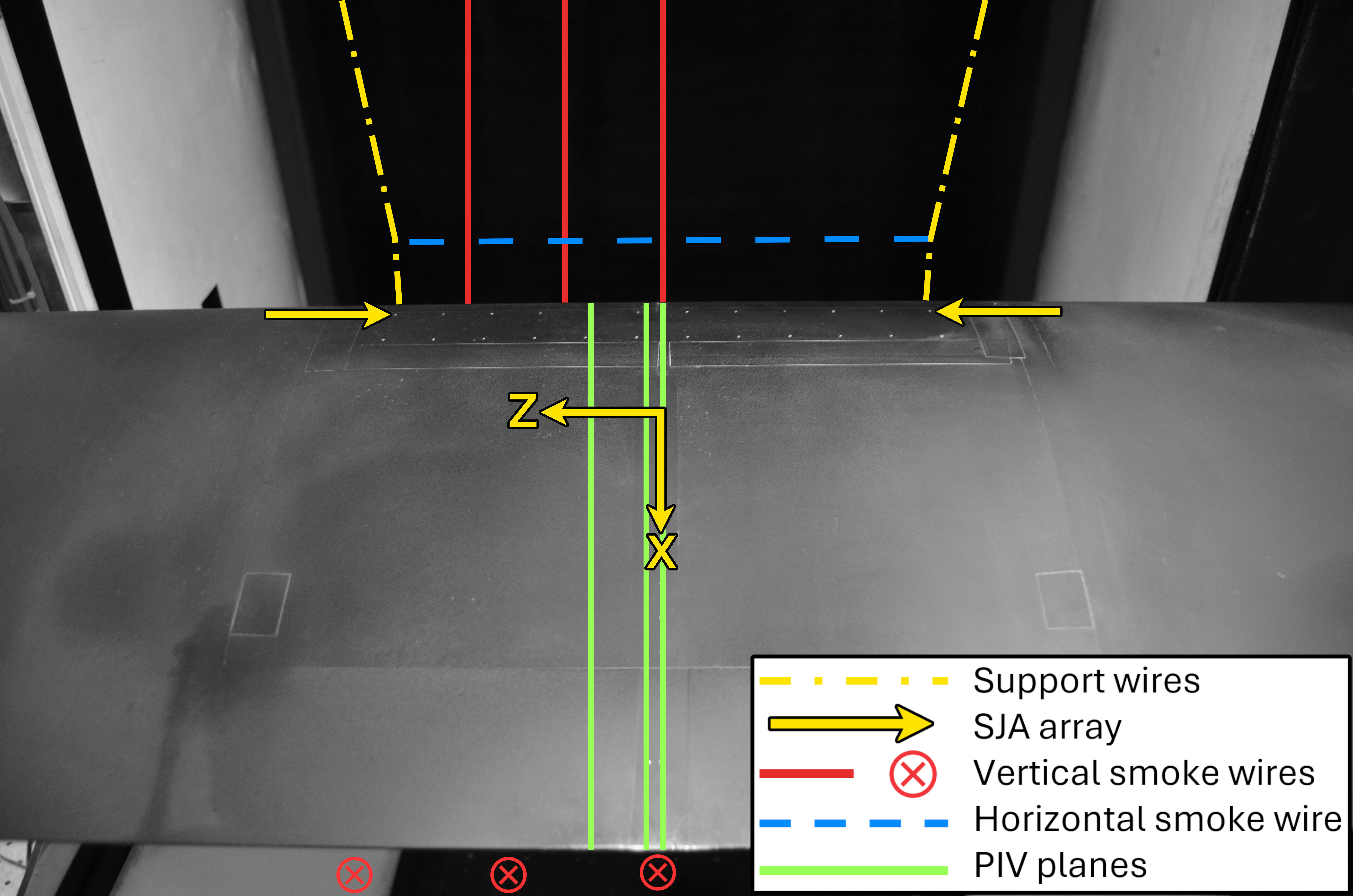}
    \caption{Labelled overhead photo of the NACA 0025 airfoil in the test section}
    \label{fig:smoke wire locations}
\end{figure}

A National Advisory Committee for Aeronautics (NACA) 0025 airfoil~\cite{Feero2015} was placed in the wind tunnel with the leading edge approximately 40~\unit{\centi\metre} from the test section inlet (Fig.~\ref{fig:smoke wire locations}). Additionally, the coordinate axes are defined, with $z=0$ corresponding to the midspan. The aluminum wing has an aspect ratio of approximately 3, with a span of $b=885$~\unit{\milli\metre}, and a chord length of $c=300$~\unit{\milli\metre}. The wing spans the entire width of the test section and features circular end plates, which isolate it from the boundary layer at the wind tunnel walls~\cite{Feero2017b}. The wing comprises three parts, with a hollow center third to house the sensors and actuators. In the center, there is a \SI{317}{\milli\metre}~$\times$~\SI{58}{\milli\metre} rectangular cutout where the microblower array is installed, with a flush 0.8 mm hole for the nozzle of each SJA. The angle of attack was set to $\alpha=\SI{10}{\degree}$, such that the flow separates at approximately 12\% chord with the specified flow parameters~\cite{Xu2023}.

\begin{figure}
    \centering
    \includegraphics[width=0.35\linewidth]{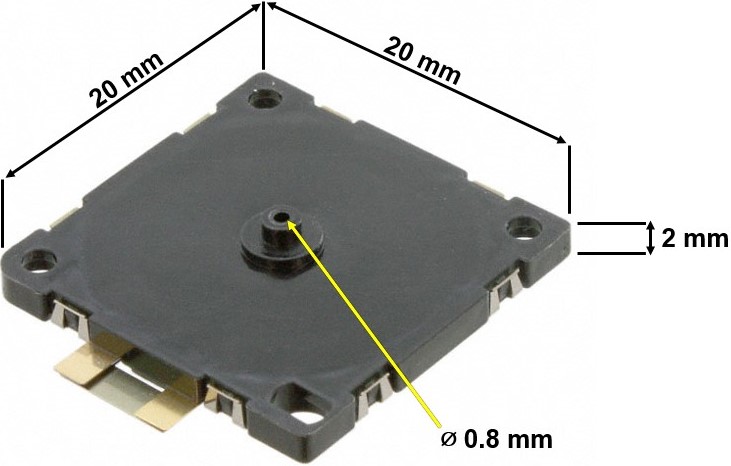}
    \caption{Murata MZB1001T02 microblower}
    \label{fig:SJA}
\end{figure}

The SJAs used are the commercially available Murata MZB1001T02 microblowers, pictured in Fig.~\ref{fig:SJA}, and are embedded underneath the surface of the wing model. The array consists of two rows of 12 SJAs located at 10.7\% and 19.8\% chord. However, only the upstream row was used in this experiment, as indicated by the arrows in Fig.~\ref{fig:smoke wire locations}. The SJA operates between 5--30~\unit{\volt} and has a drive resonant frequency between 24--27~\unit{\kilo\hertz}. The mean centerline velocity of the synthetic jet reached a maximum when driven at a frequency of \SI{25.1}{\kilo\hertz}. However, due to the right-skewed velocity response, the carrier frequency was chosen as $f_c=25.5$~\unit{\kilo\hertz} to ensure a stable jet velocity~\cite{Xu2023}. In this investigation, the SJAs were modulated at two excitation frequencies $f_m=20$~\unit{\hertz} and $200$~\unit{\hertz}, corresponding to non-dimensional frequencies of $F^+=1.18$ and $11.76$, respectively. Square waveforms were used for the carrier and modulation frequency, with a duty cycle of 50\%. The blowing strength of SJAs is often quantified by the blowing ratio, $C_B=\overline{U_j}/U_\infty$, where $\overline{U_j}$ is the time-averaged jet velocity. However, for an array of discrete SJAs, the blowing strength is more appropriately represented by the momentum coefficient~\cite{Amitay2001}, \begin{equation}C_\mu=\frac{\overline{I_j}}{\frac{1}{2}\rho_oA_fU_\infty^2}\end{equation}
where $\overline{I_j}$ is the time-averaged jet momentum, $\rho_o$ is the freestream fluid density, and $A_f$ is the projected control area. For both control cases in this study, the SJAs were operated at 20 $\mathrm{V}_{pp}$, corresponding to a momentum coefficient of $C_\mu=2.0\times10^{-3}$.

\begin{figure}
    \centering
    \includegraphics[width=0.5\linewidth]{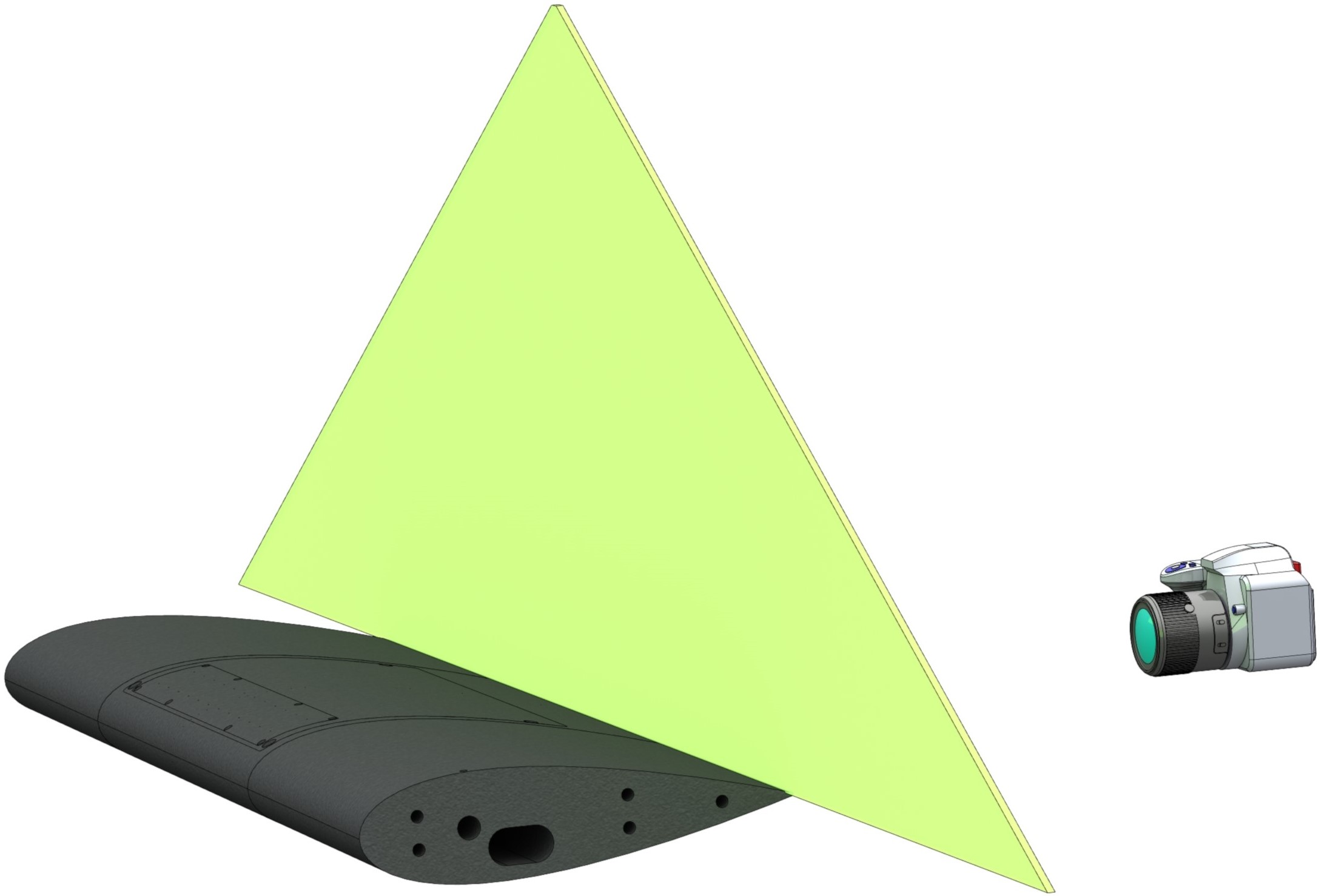}
    \caption{Depiction of the laser sheet and camera orientation for the sectional smoke flow visualization}
    \label{fig:yz smoke}
\end{figure}

Smoke flow visualization was performed with 1) upstream and downstream vertical smoke wires indicated in red in Fig.~\ref{fig:smoke wire locations}; and 2) a single upstream horizontal smoke wire (blue dashed line)~\cite{Machado2024}. The vertical smoke wires were kept taut with weights attached to the bottom of the wires underneath the wind tunnel. The horizontal smoke wire was installed upstream of the wing, along the chord line, allowing for visualization of the flow at the edge of the shear layer. The horizontal smoke wire was kept taut by making it slightly shorter than the distance between the support wires (yellow dash-dotted lines) so that they bowed inward and applied tension. A Nikon D7000 DSLR camera was used to image the smoke visualization. For the vertical smoke wire configurations, the camera captured the side view of the airfoil from outside the test section. The trailing edge visualization of the $z$-$y$ plane was imaged with the camera downstream of the airfoil and test section. Lastly, the overhead visualization was imaged with the camera atop the wind tunnel through an acrylic observation window. The smoke streaks were illuminated with a Nikon SB-800 speedlight or a continuous laser set atop the wind tunnel. The speedlight provided a quick burst of illumination, which limited the exposure time to 1/5900 seconds allowing for the capture of small-scale flow structures. Alternatively, the laser sheet provided two-dimensional illumination, allowing for a cross-sectional view of the flow at the trailing edge (Fig.~\ref{fig:yz smoke}). Additionally, the continuous laser sheet allowed for longer exposure times, allowing for visualization of the mean flow. The camera settings are summarized in Table~\ref{tab:camera_settings}.

\begin{table}
\caption{\label{tab:camera_settings} Summary of camera settings for smoke visualization}
\begin{ruledtabular}
\begin{tabular}{lccccr}
Visualization plane&Smoke wire&Exposure time (\unit{\second})&ISO&Aperture&Illumination\\
\hline
streamwise-transverse ($x$-$y$) & vertical & 1/5900 & 100 & $f/$8.0 & speedlight\\
spanwise-transverse ($z$-$y$) & horizontal & 1/6 & 25600 & $f/$1.8 & laser sheet\\
streamwise-spanwise ($x$-$z$) & horizontal & 1/5900 & 640 & $f/$8.0 & speedlight\\
\end{tabular}
\end{ruledtabular}
\end{table}

The airfoil surface pressure was measured with an MKS Baratron 226A differential pressure transducer, with a bidirectional range of $\pm26.7$~\unit{\pascal}. This measurement was conducted in conjunction with a Scanivalve pressure scanner, connected to pressure taps located along the midspan of the airfoil. The dynamic pressure of the freestream was measured with a pitot-static tube connected to a separate pressure transducer with a range of 267~\unit{\pascal}. For each measurement, 30,000 samples were collected at a sampling rate of \SI{1}{\kilo\hertz}. In our analysis, we focused on pressure readings from a single tap, located at $x/c=0.8$. The pressure coefficient, $C_p$, was calculated as
\begin{equation}
    C_p=\frac{p-p_\infty}{p_0-p_\infty}
\end{equation}
where $p-p_\infty$ is the measured pressure difference between the airfoil pressure and the static pressure, and $p_0-p_\infty$ is the measured pressure difference between the stagnation pressure and the static pressure.

Time-resolved velocity measurements were achieved using hot-wire anemometry. Measurements were taken using a DANTEC 55C17 CTA bridge, paired with a 55P01 single-wire probe of \SI{5}{\micro\metre} diameter and \SI{1}{\milli\metre} sensing length. To enhance measurement accuracy, a high overheat ratio of 1.6 was used. Data were sampled at a rate of \SI{20}{\kilo\hertz}, and $2.4\times10^6$ samples were collected for each measurement location. Before measurement, the hot-wire was calibrated against a pitot tube using King’s Law. The hot-wire was attached to a long, thin aluminum probe holder (\SI{234}{\milli\metre} long and \SI{4}{\milli\metre} diameter) to minimize the invasiveness of the measurement. The position of the hot-wire probe was computer-controlled with a 3-axis traverse system equipped with stepper motors, ensuring accurate and repeatable motion with fine resolutions ($<\SI{0.4}{\milli\metre\per step}$).

\begin{figure}
    \centering
    \includegraphics[width=0.5\linewidth]{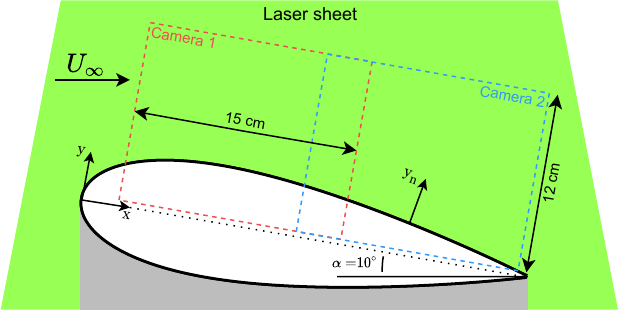}
    \caption{Camera FOVs and laser sheet orientation for the PIV experiment}
    \label{fig:pivsetup}
\end{figure}

Particle image velocimetry (PIV) measured the streamwise-transverse velocity field above the airfoil. Neutrally buoyant particles were introduced into the flow using a fog machine situated downstream of the test section. Two JAI SP500-USB cameras, each with a resolution of 2560~$\times$~2048, were positioned outside the test section and aligned with the airfoil's chord. Composite images were created by stitching together the camera captures, resulting in a field of view (FOV) measuring \SI{270}{\milli\metre}~$\times$~\SI{120}{\milli\metre}, with a pixel resolution of \SI{17}{pixels\per\milli\metre} (Fig.~\ref{fig:pivsetup}). The stitched FOV effectively captured the range $x/c \in [0.1,1]$ and $y/c \in [0,0.4]$, covering the area of interest above the airfoil. A Litron Bernoulli \SI{200}{\milli\joule} Nd-YAG laser with a wavelength of \SI{532}{\nano\metre} was passed through converging and diverging cylindrical lenses ($f=1000$~\unit{\milli\metre} and $f=-13.7$~\unit{\milli\metre}, respectively) to create a thin laser sheet that illuminated the measurement plane. The image acquisition and laser pulses were synchronized using an NI PCI-6232e data acquisition card (DAQ) at \SI{10}{\hertz}, and 1000 image pairs were recorded for each measurement. The time delay between the two frames in an image pair was set to $\Delta=\SI{120}{\micro\second}$. The random error in the mean velocity was found to be less than $0.01U_\infty$ based on a convergence study. For the control cases, phase-locked velocity measurements were achieved by synchronizing image acquisition at 8 evenly spaced phases relative to the SJA modulation frequency, where a phase angle of $\phi=0$\si{\degree} represents the activation of the actuator. Coherent fluctuations are extracted from the velocity signal using the triple decomposition~\cite{Hussain1970}. The streamwise velocity was decomposed as $u=\bar{u}+\tilde{u}+u'$, where $\bar{u}$ is the time-averaged velocity (from 8000 samples over all 8 phase angles), $\tilde{u}$ is the coherent velocity obtained from the phase-average, and $u'$ is the fluctuation velocity~\cite{Buchmann2013}. To facilitate PIV measurements at various spanwise planes, the laser optics were mounted to a motorized linear positioning system, enabling movement of the light sheet in the $z$-direction. A 200-step-per-revolution stepper motor controlled the linear motion, translating each revolution into 1 mm of precise and repeatable movement. Velocity vectors were extracted from the particle images using a multi-pass cross-correlation to accurately record both large and small displacements~\cite{Thielicke2021}. An initial interrogation window of 64~$\times$~64 pixels was used, followed by two passes with the window length decreasing by a factor of 2 each pass. Linear window deformation was used to reduce correlation errors in high shear regions.

The instantaneous boundary layer thickness ($\delta_I$) and its distribution are used as an indicator of stability. Specifically, we aim to find the distribution of wall-normal distances where the streamwise velocity surpasses $0.99U_{\infty}$. To extract meaningful information from experimental noise, a series of velocities at points forming a line tangential to the wall are averaged to estimate the velocity at a given wall-normal distance. Averaging is only done in the wall-tangential direction, and not the wall-normal direction, to ensure a high $y_n$ resolution and a meaningful distribution. The accuracy of the averaging method can be evaluated from $\epsilon=\bar{\delta_I}-\delta_m$ where $\bar{\delta_I}$ is the mean of the instantaneous boundary layers, and $\delta_m$  is the boundary layer thickness estimated from the time-averaged velocity field. For the present case, the wall-normal velocity was obtained by averaging the velocity along a line of length $0.05c$. The error between instantaneous and mean estimates was found to be $\epsilon<5\%$ and $\epsilon<10\%$ for $z/c=0.03$ and $z/c=0.12$, respectively. This low error figure validates that the average of the instantaneous boundary layer thicknesses serves as a meaningful estimator of the mean boundary layer thickness.

\section{Results}
\subsection{Influence of Actuation Frequency on Aerodynamic Stability}
\label{frequency study}
\begin{figure*}
    \begin{subfigure}{0.325\linewidth}
        \includegraphics[width=\linewidth]{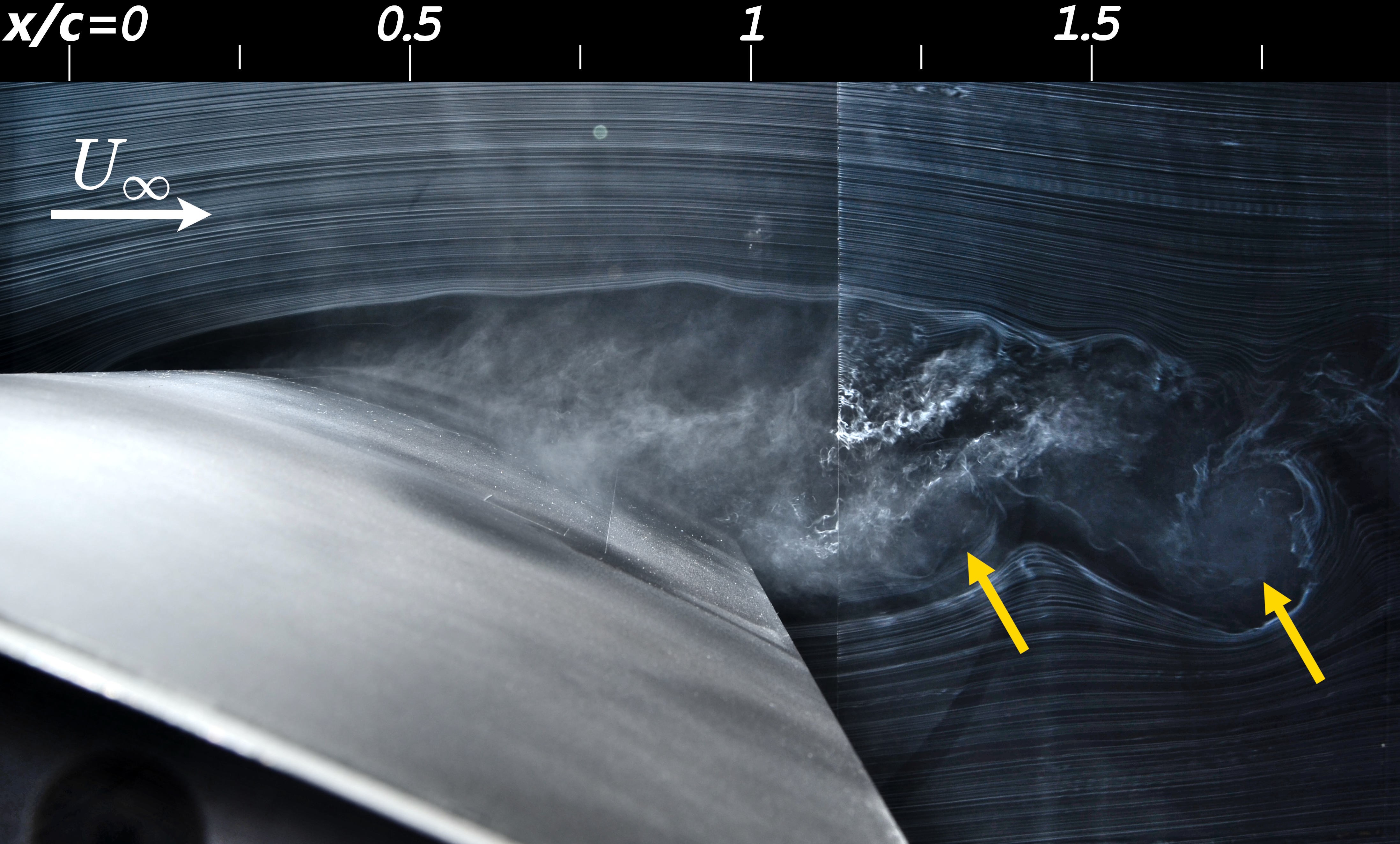}
        \caption{Baseline}
        \label{fig:smoke_baseline}
    \end{subfigure}
    \begin{subfigure}{0.325\linewidth}
        \includegraphics[width=\linewidth]{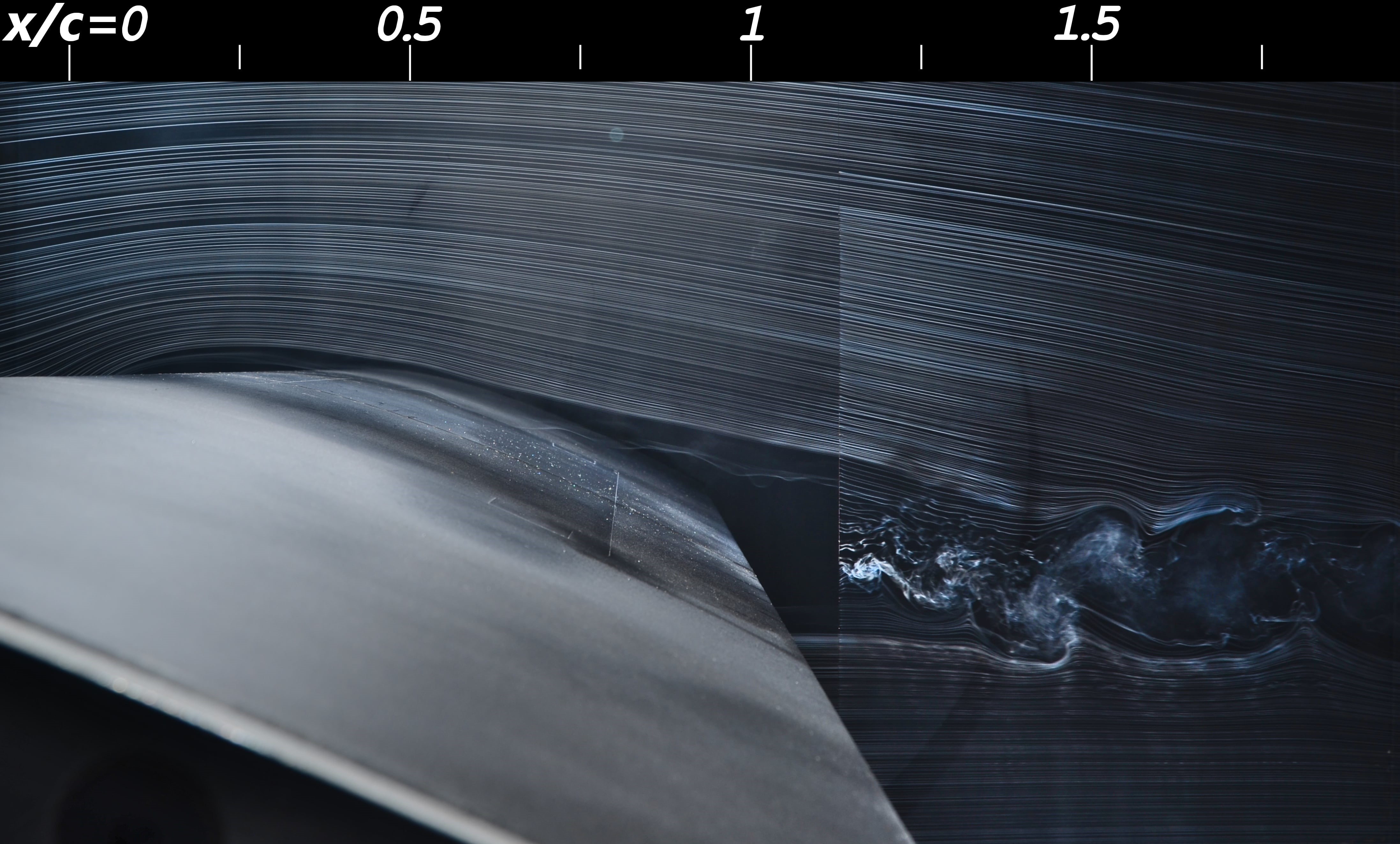}
        \caption{$F^+=1.18$}
        \label{fig:smoke_F1}
    \end{subfigure}
    \begin{subfigure}{0.325\linewidth}
        \includegraphics[width=\linewidth]{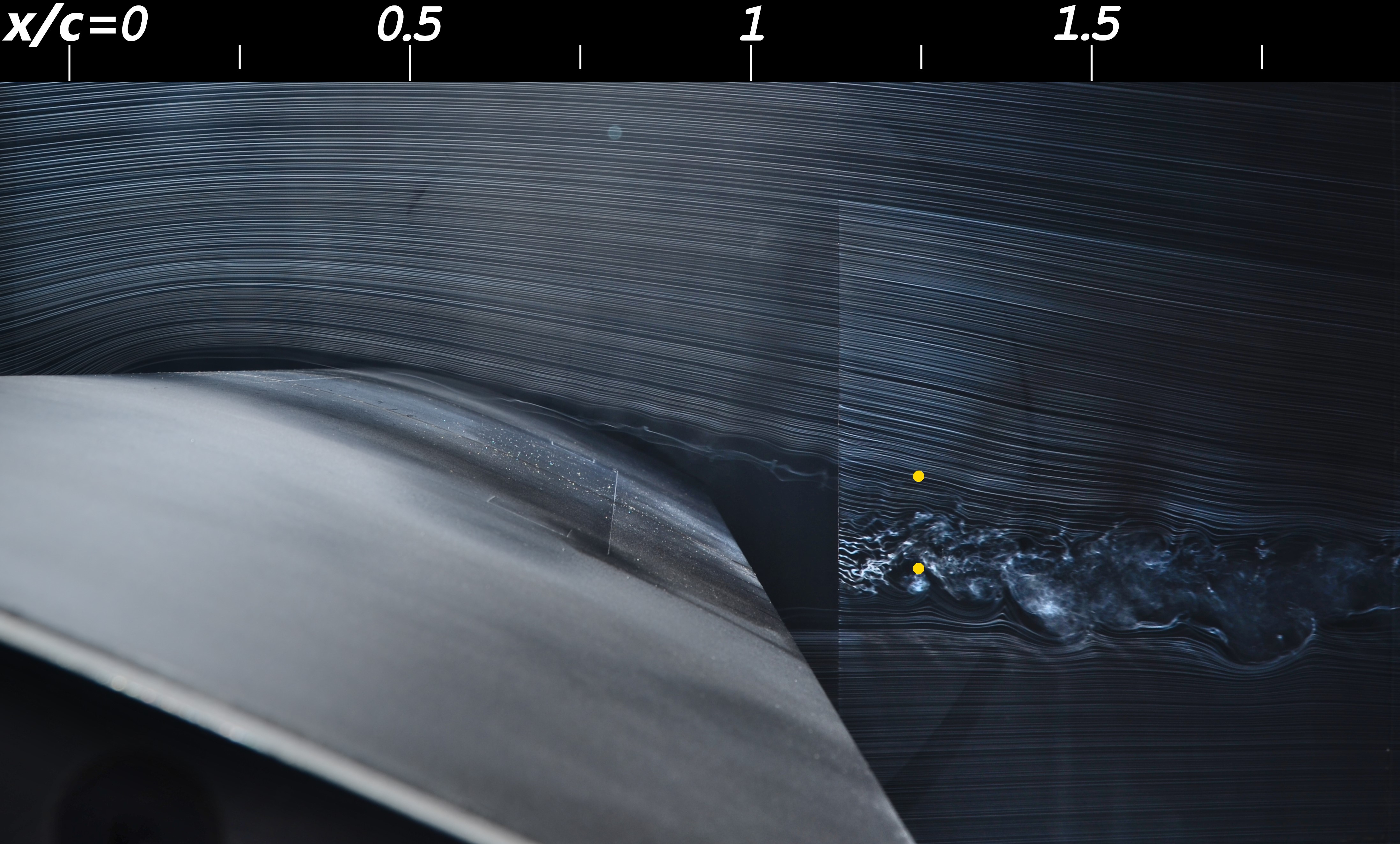}
        \caption{$F^+=11.76$}
        \label{fig:smoke_F11}
    \end{subfigure}
    \caption{Smoke flow visualization at the midspan; streamwise-transverse plane}
    \label{fig:smoke}
\end{figure*}

Smoke flow visualizations of the baseline flow, and the two control cases are presented in Fig.~\ref{fig:smoke} for the $x$-$y$ plane. In the baseline case, flow separation occurs with a large recirculation area revealed by the upstream motion of smoke generated from the downstream wire, and the high trajectory of the laminar streaklines above. As expected with a stalled flow, a wide wake accompanied by large-scale vortex shedding is observed, as highlighted by the yellow arrows. The visualizations reveal that both control frequencies result in complete flow reattachment, as evident by the absence of reverse flow observed from the downstream smoke wire, however, clear differences can be seen in the wake. Low-frequency control at $F^+=1.18$ results in an alternating vortex street, with large-scale vortical structures dominating the wake dynamics. In contrast, the high-frequency control ($F^+=11.76$) results in finer structures in the wake compared to that of the low-frequency control, marked by the absence of large-scale vortex shedding and a narrower wake profile, see~Machado et al.~\cite{Machado2024}. Lastly, a wavy pattern appears in the first few laminar streaklines above the wing surface and extending downstream of the trailing edge, exclusively in the high-frequency control case. This pattern suggests that flow structures induced by high-frequency actuation remain coherent downstream of the airfoil, potentially interacting with the shear layer, while remaining decoupled from the shedding frequency. To further investigate this flow phenomenon, we probe this specific location with a hot-wire to obtain time-resolved velocity data. The hot-wire measurement locations are shown as yellow dots in Fig.~\ref{fig:smoke_F11}.

\begin{figure}
    \centering
    \begin{subfigure}{\linewidth}
            \includegraphics[width=0.5\linewidth]{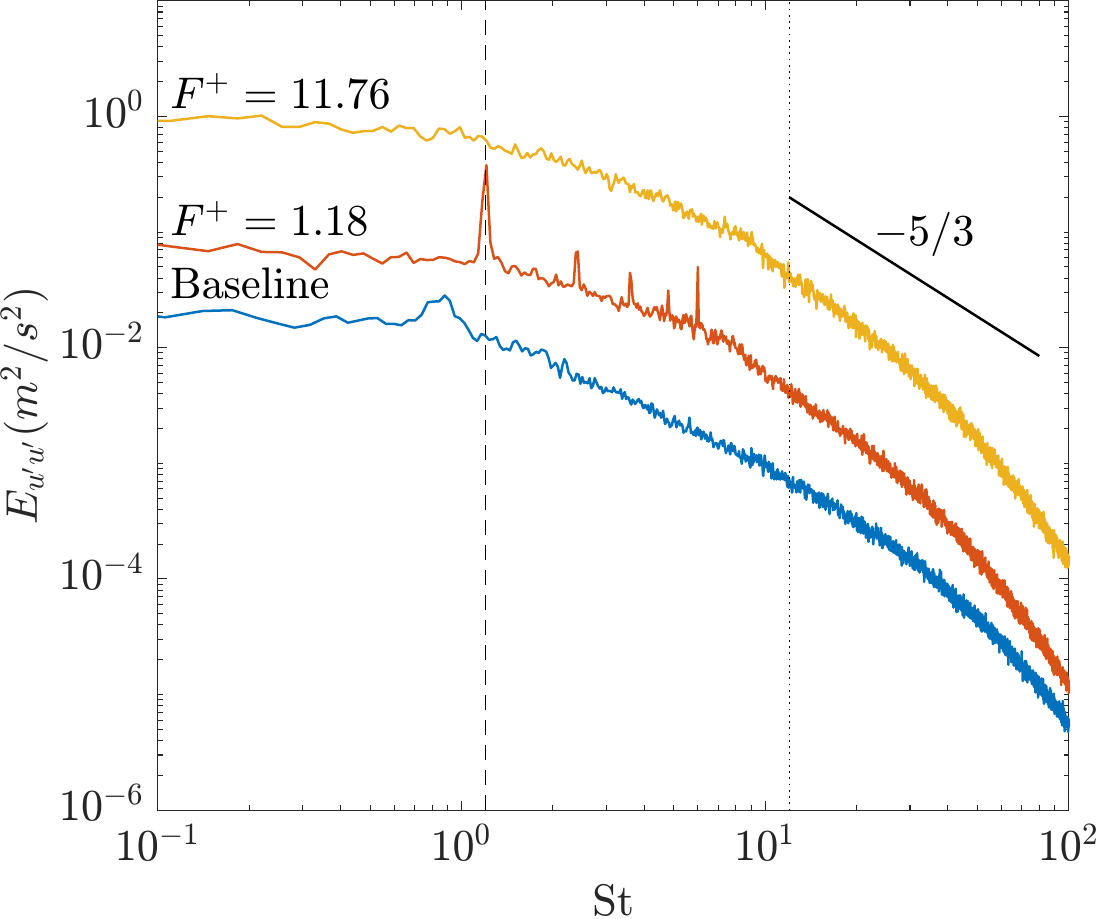}
    \caption{Mid-wake ($y/c=0$)}
    \label{fig:spectra mid-wake}
    \end{subfigure}
    \begin{subfigure}{\linewidth}
            \includegraphics[width=0.5\linewidth]{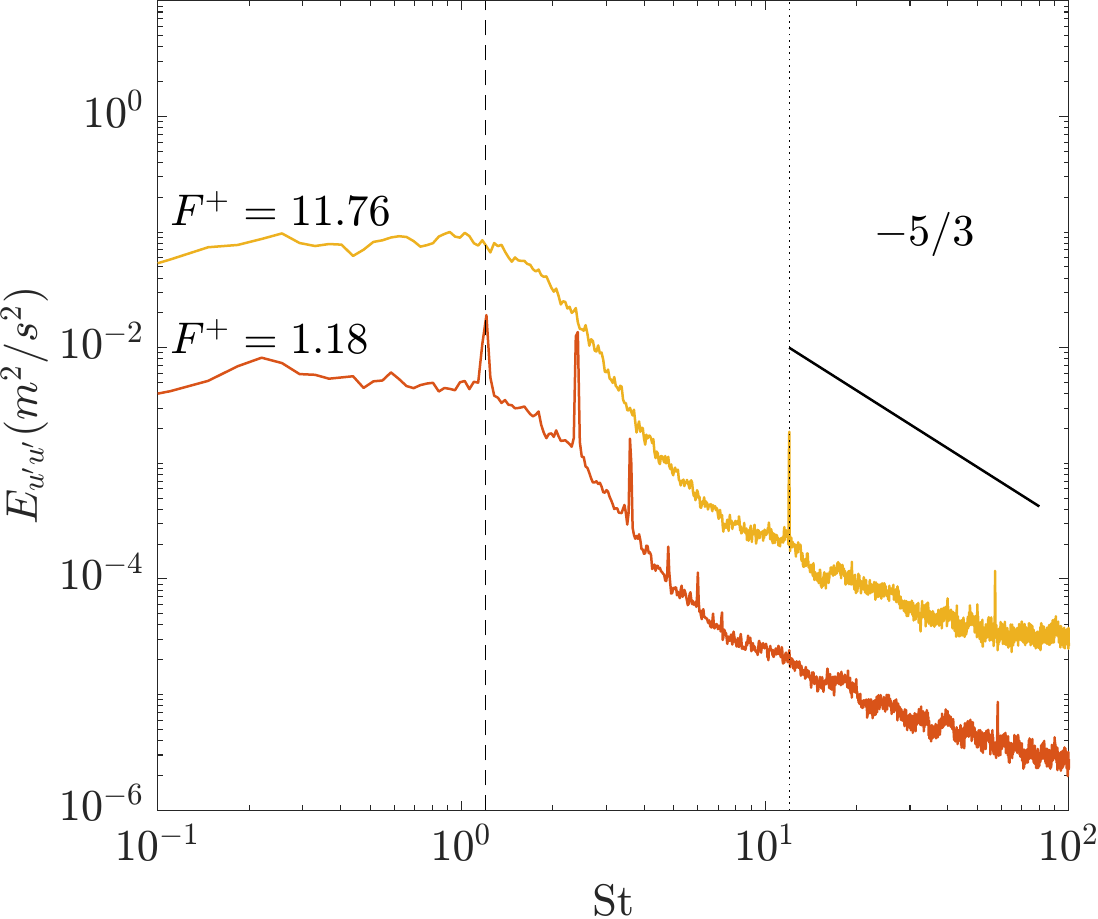}
    \caption{Shear layer boundary ($y/c=0.14$)}
    \label{fig:spectra shear layer}
    \end{subfigure}
    \caption{Streamwise velocity spectra of the wake at $x/c=1.25$. Subsequent spectra are stepped by one decade for clarity. $\mathrm{St}=1.18$ is highlighted by a dashed line, and $\mathrm{St}=11.76$ a dotted line}
    \label{fig:spectra}
\end{figure}

Streamwise velocity spectra of the near wake ($x/c=1.25$) are presented in Fig.~\ref{fig:spectra} for the baseline case, and both control cases. The mid-wake spectra of the baseline flow (Fig.~\ref{fig:spectra mid-wake}) exhibits a broadband peak, revealing that the wake shedding frequency is centered around $\mathrm{St_w=0.85}$. This result is in agreement with past work~\cite{Feero2015, Xu2023} in which the same experimental facilities and flow parameters were used. With control at $F^+=1.18$, a sharp spectral peak appears at the actuation frequency, indicating that the forcing frequency is driving the highly organized vortex shedding in the wake~\cite{Glezer2005,Feero2015,Xu2023}. This periodicity results in a wider and less steady wake. Conversely, high-frequency control at $F^+=11.76$ suppresses the natural shedding frequency of the airfoil, as evidenced by the absence of a spectral peak at $\mathrm{St}\approx\mathcal{O}(1)$. Instead, a turbulent spectral profile is observed, resulting in a more uniform, steady, and consequently narrower wake, as seen in Fig.~\ref{fig:smoke_F11}. To investigate the persistence of flow structures downstream of the airfoil's trailing edge, a hotwire survey was conducted across the entire width of the wake. The velocity spectra across the wake was analyzed at \SI{1}{\centi\metre} intervals, i.e. $\mathrm{d}y=c/30$. Under control at $F^+=11.76$, only two locations near the edge of the shear layer exhibited a sharp peak corresponding to the actuation frequency. The velocity spectra containing the strongest peak was measured at $y/c=0.14$ and is plotted in Fig.~\ref{fig:spectra shear layer}. The occurrence of a spectral peak at this precise location confirms the presence of persistent, small-scale structures that are induced by the high-frequency actuation of the SJAs. It is hypothesized that these structures help dissipate large-scale vortices in the wake, resulting in the smooth controlled streaklines above the wake, as seen in Fig.~\ref{fig:smoke_F11}. The visualization of the wake dynamics in conjunction with the lack of periodic spectral components in the mid-wake suggests that high-frequency control results in a steadier and lesser pressure drag force compared to low-frequency actuation.

\begin{figure}
    \centering
    \includegraphics[width=0.5\linewidth]{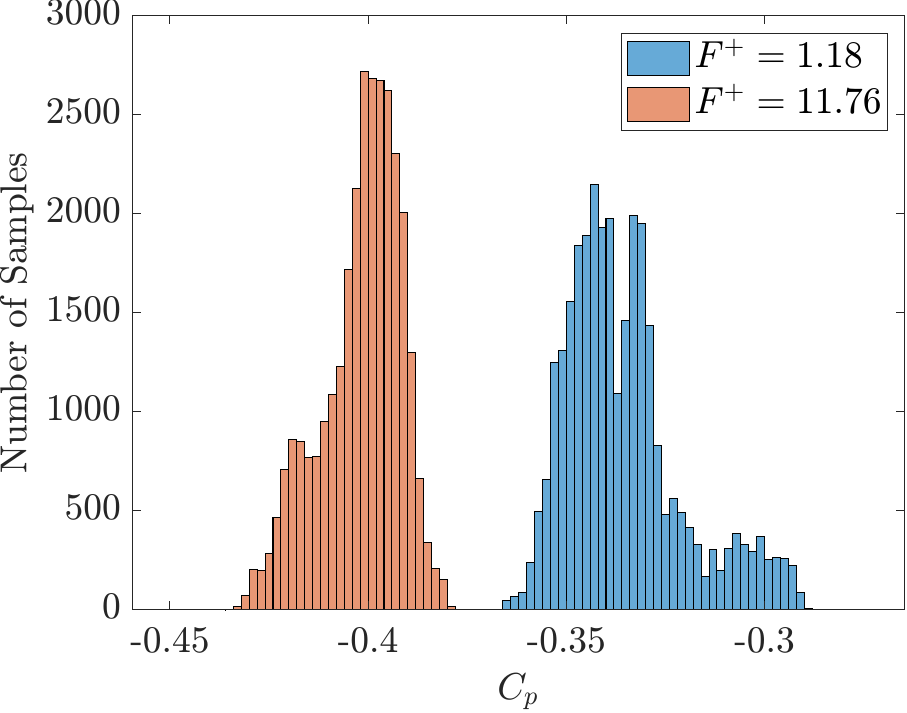}
    \caption{Distribution of surface pressure coefficients at $x/c=0.8$}
    \label{fig:pressure histogram}
\end{figure}

To investigate the stability of the shear layer, a histogram of surface pressure measurements near the trailing edge is plotted in Fig.~\ref{fig:pressure histogram}. Firstly, it is evident that the high-frequency control case results in greater suction pressure at this chordwise location. However, it should be noted that modulation at $F^+=1.18$ results in a higher time-averaged lift coefficient, due to a larger suction peak upstream~\cite{Xu2023}. The surface pressure distributions highlight a stark contrast between the two control cases, with the high-frequency control yielding a much narrower distribution. The narrower distribution of the high-frequency control case indicates a much steadier boundary layer, and thus a more stable lift force. The low-frequency control results in a bimodal distribution. Previous results~\cite{Machado2024} showed a spanwise contraction in the flow towards the midspan at both control frequencies. With actuation at $F^+=1.18$, the chordwise position of this contraction exhibited temporal variability, and was attributed to the convection of large spanwise vortex rollers induced by the SJAs. Conversely, control at $F^+=11.76$ resulted in a more gradual and time-invariant flow contraction. The higher pressure peak associated with less samples is likely due to a spanwise vortex being present above the pressure tap at the time of measurement. This shows that large vortices induced by low-frequency actuation results in an unsteady boundary layer and a time-varying lift force.

\begin{figure*}
    \begin{subfigure}{0.325\linewidth}
        \includegraphics[width=\linewidth]{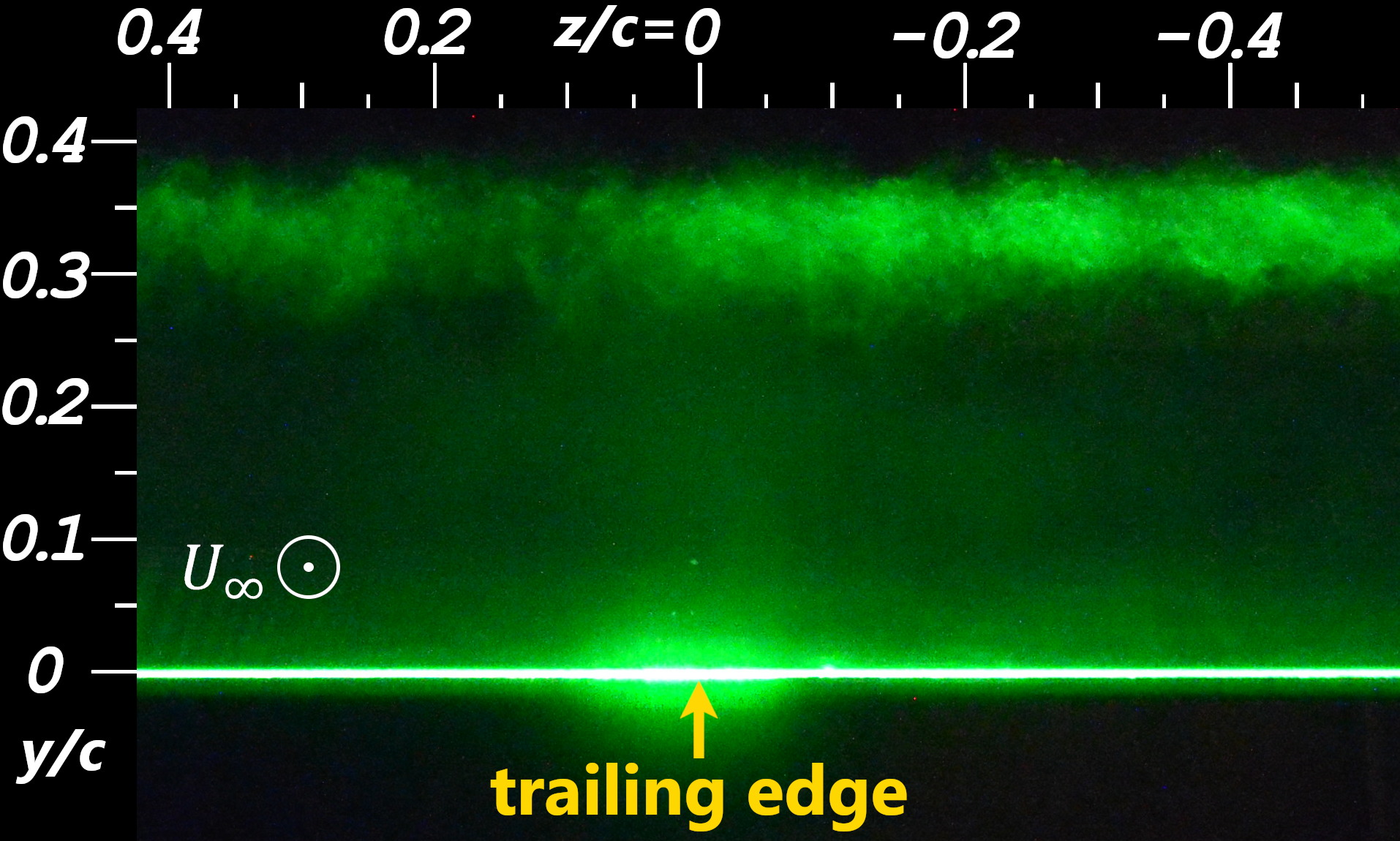}
        \caption{Baseline}
        \label{fig:laser_smoke_baseline}
    \end{subfigure}
    \begin{subfigure}{0.325\linewidth}
        \includegraphics[width=\linewidth]{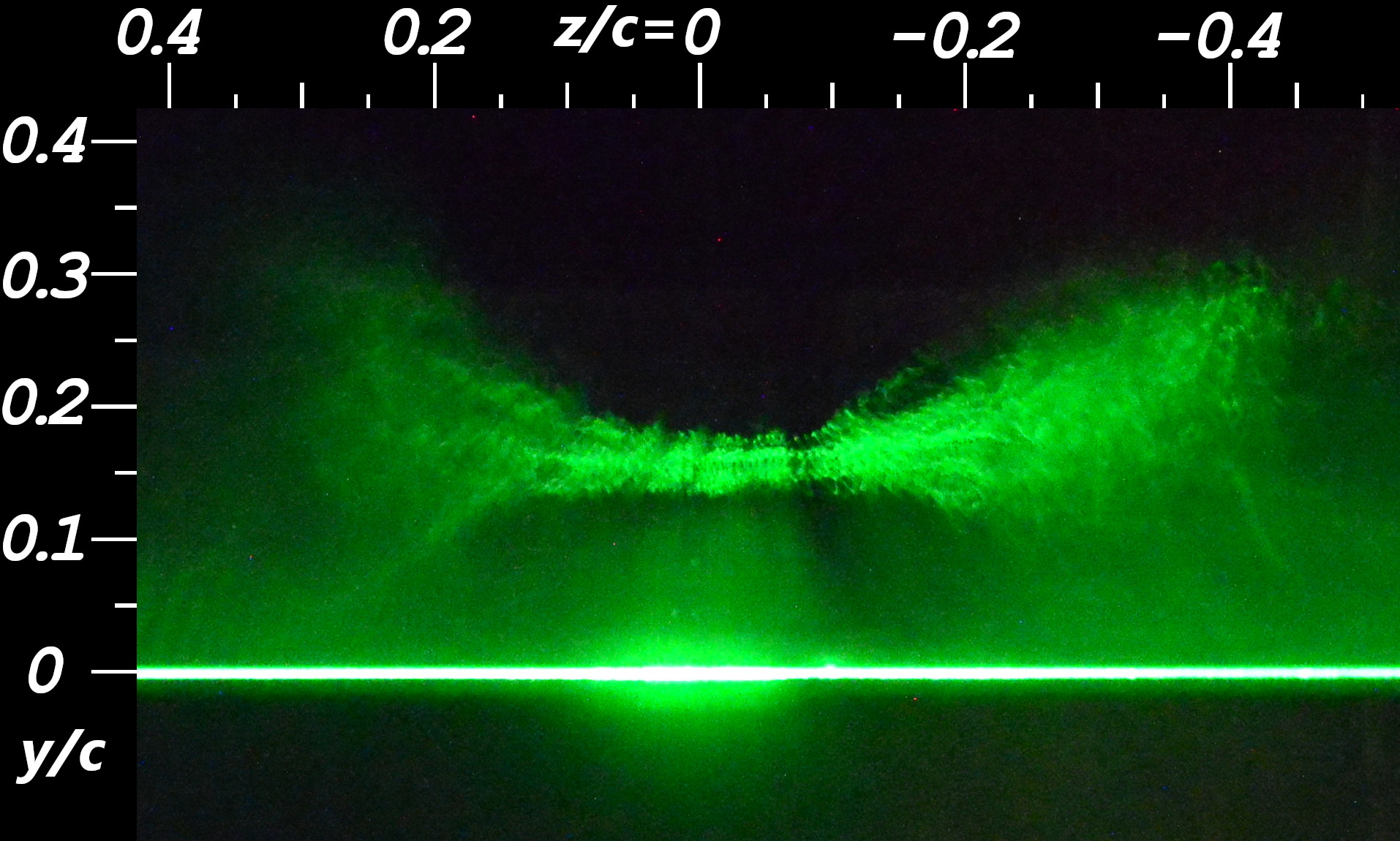}
        \caption{$F^+=1.18$}
        \label{fig:laser_smoke_F1}
    \end{subfigure}
    \begin{subfigure}{0.325\linewidth}
        \includegraphics[width=\linewidth]{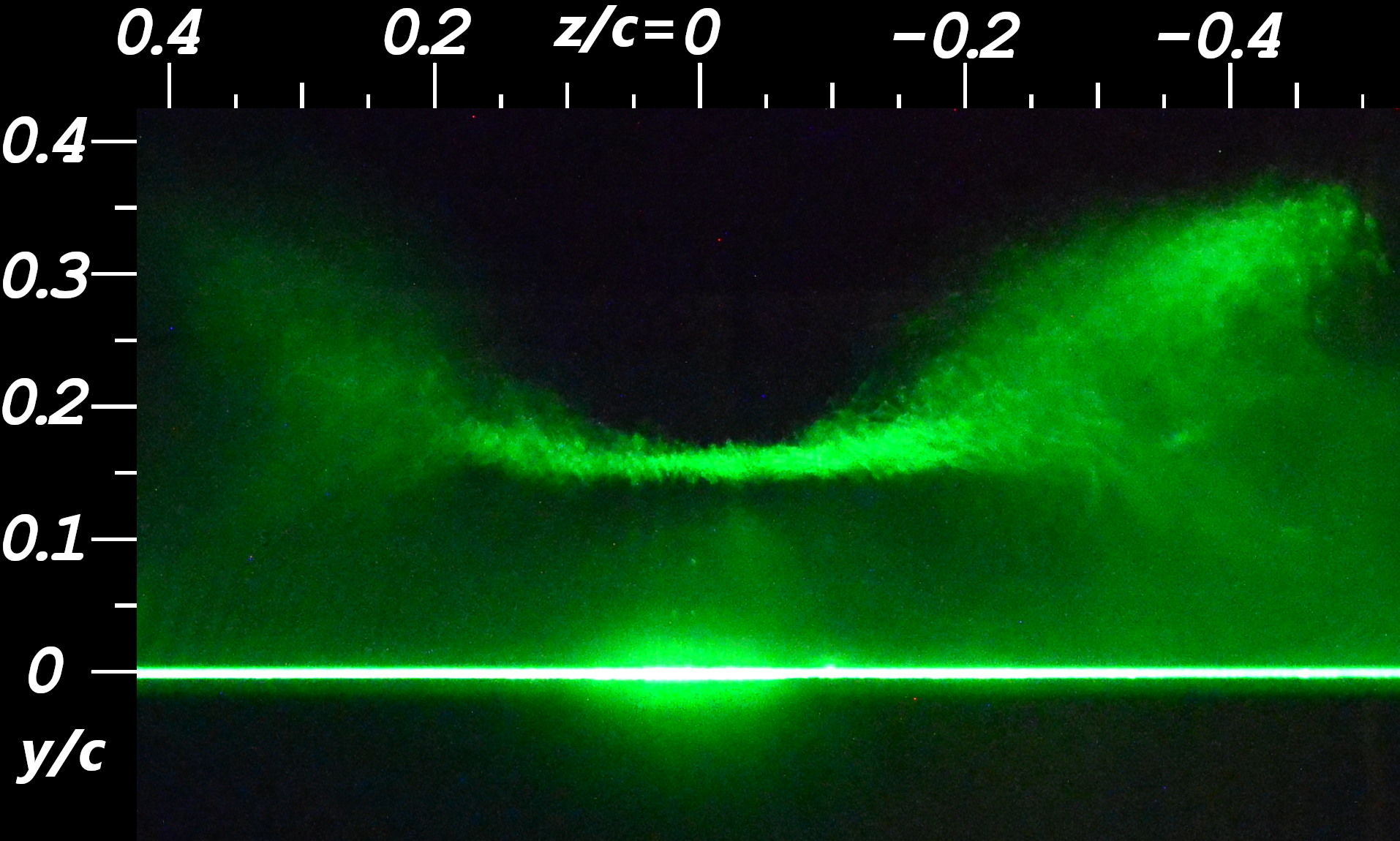}
        \caption{$F^+=11.76$}
        \label{fig:laser_smoke_F11}
    \end{subfigure}
    \caption{Sectional smoke flow visualization at the trailing edge; spanwise-transverse plane}
    \label{fig:laser_smoke}
\end{figure*}

To further investigate the stability of the controlled flow at the two actuation frequencies, sectional flow visualizations at the trailing edge are presented in Fig.~\ref{fig:laser_smoke}. The illuminated smoke in the images highlights the shear layer boundary across the span. In the baseline case (Fig.~\ref{fig:laser_smoke_baseline}), the boundary layer at the trailing edge is large and highly unsteady as evidenced by the bar of diffuse smoke centered at $y/c=0.35$. The visualizations of the control cases reveal an effective control length of approximately 0.4$c$, centered about the midspan, where the streaklines are dense and near to the airfoil surface. However, beyond $\pm$0.2$c$, diffuse smoke is observed, extending higher above the airfoil surface, indicating unsteady shear layer behavior and the breakdown of flow control. These results are in agreement with previous flow visualizations~\cite{Machado2024} which showed that the control breaks down between $z/c=0.17$ and $z/c=0.33$. The spanwise control authority of the SJA array is discussed further in Section~\ref{sec:spanwise}. In the low-frequency control case (Fig.~\ref{fig:laser_smoke_F1}), a somewhat unsteady shear layer is observed as the smoke varies in the $y$ position during the exposure of the image. In contrast, a much tighter spread of smoke is observed for the high-frequency control case (Fig.~\ref{fig:laser_smoke_F11}), indicating that the shear layer thickness is more constant. These differences are most prominent at the edges of the effective control region at $\pm$0.2$c$. This result suggests that high-frequency actuation results in a steadier shear layer, and thus a steadier lift force, which is in agreement with the analysis of the surface pressure distributions.

\subsection{High-Frequency Control: Vortex Ring Dynamics}
\label{VRs}
\begin{figure*}
    \centering
    \includegraphics[width=\linewidth]{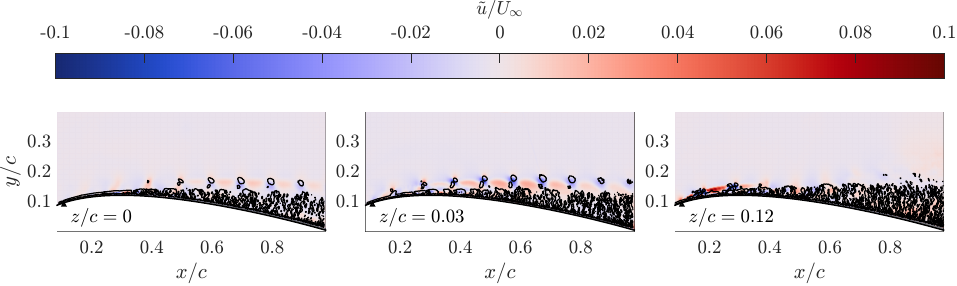}
    \caption{Contours of coherent streamwise velocity with iso-contour of $Q=300$ overlaid at multiple spanwise planes. $\phi=0$\si{\degree}, $F^+=11.76$}
    \label{fig:Qcrit}
\end{figure*}

The $Q$-criterion~\cite{Hunt1998} was used to identify vortices in the controlled flowfield. $Q$ is defined as the second invariant of the velocity gradient tensor, and can be expressed as
\begin{equation}
    Q=\frac{1}{2}(|\omega_z|^2-S^2)
\end{equation}
for the 2D velocity field, where $\omega_z$ is the out-of-plane vorticity, and $S$ is the strain rate tensor. Iso-contours of $Q=300$ are plotted in Fig.~\ref{fig:Qcrit} for two closely spaced measurement planes: midspan and at $z/c=0.03$, as well as at a plane farther from the midspan, $z/c=0.12$. Additionally, to help discern the rotational sense of identified vortices, contours of the coherent streamwise velocity ($\tilde{u}$) are overlaid. The chordwise location of the SJA is denoted by the black triangle at $x/c=0.1$. Small-scale vortex pairs are identified above the boundary layer in both spanwise planes near the midspan. In between the two vortices exists a concentrated area of negative coherent streamwise velocity, whereas above and below the vortex pair, positive velocities are observed. From this, we discern that the top vortex is rotating clockwise, and the bottom, counter-clockwise, together appearing as a counter-rotating vortex pair in the 2D measurement plane. The absence of coherent structures in the third plane will be discussed in Section~\ref{sec:spanwise}.

\begin{figure}
    \centering
    \includegraphics[width=0.5\linewidth]{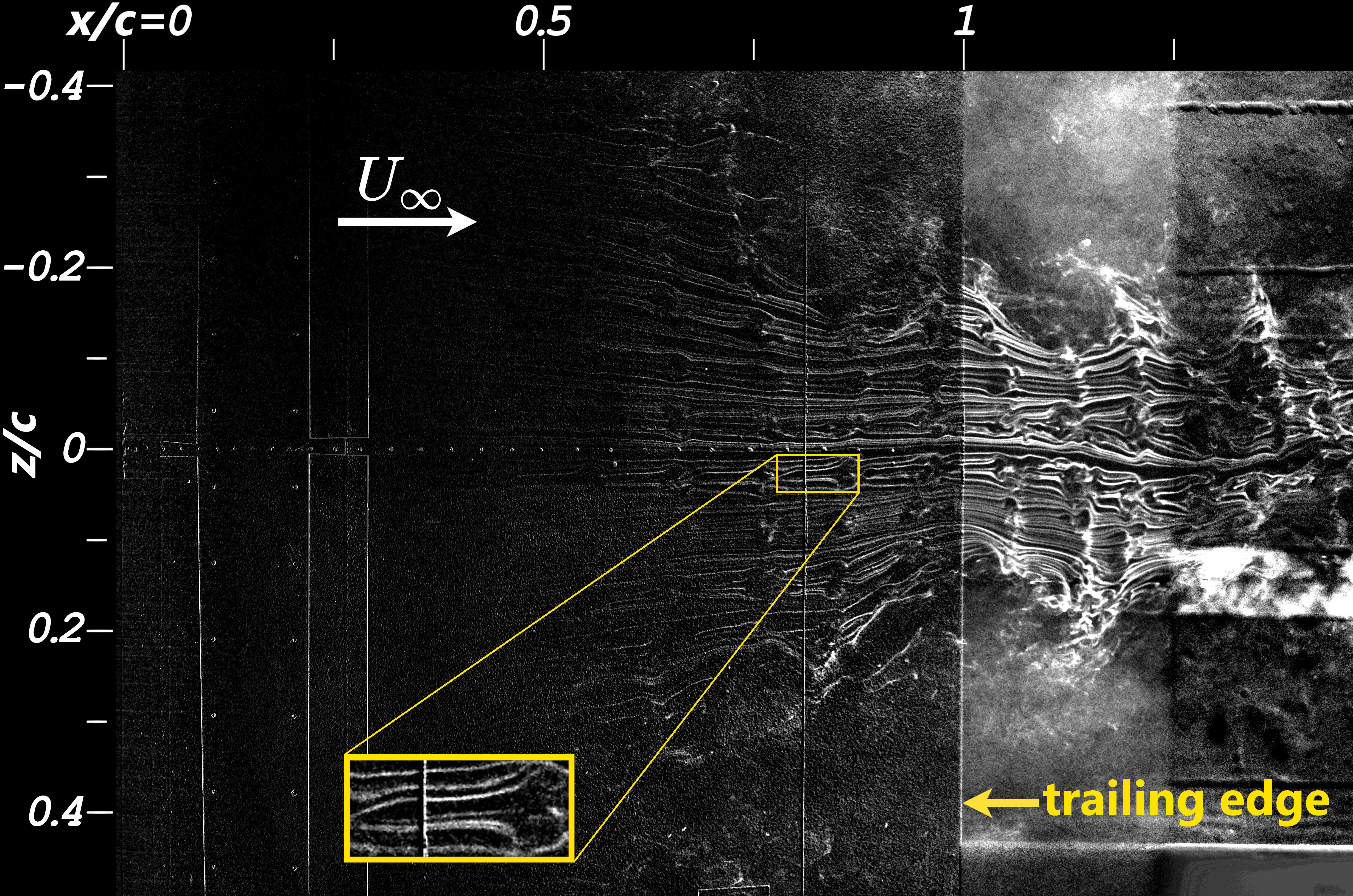}
    \caption{Overhead smoke visualization of flow above the airfoil, $F^+=11.76$}
    \label{fig:smoke_overhead}
\end{figure}

Complementing this sectional view of the flow structures, an overhead smoke visualization is presented in Fig.~\ref{fig:smoke_overhead}. The streaklines follow the flow at the edge of the boundary layer, providing visualization of the flow around the aforementioned structures in the streamwise-spanwise ($x$-$z$) plane. The patterns observed in the streaklines are spaced 0.1$c$ apart in the streamwise direction, which is consistent with the spacing between the vortices identified by the $Q$-criterion in Fig.~\ref{fig:Qcrit}. The yellow box contains a magnified view of the streaklines around the flow structure, in which they are seen to diverge then curl inward and back towards the center --- a flow pattern that is also consistent with counter-rotating vortices. Unifying the orthogonal perspectives provided by these two figures, we conclude that the observed structures are indeed vortex rings (VRs) created by the SJA each cycle, with their toroidal axes aligned with the flow direction. The presence and persistence of the observed VRs align with previous studies~\cite{Gordon2002,Zhong2005,Jabbal2008, Ho2022} which also documented the formation and advection of SJA induced VRs in crossflows. The studies demonstrate that SJAs create VRs once a threshold blowing ratio is met; below this threshold, asymmetrical vortical structures form instead. In the present study the blowing ratio of the SJA ($C_B=4.8$) is much greater than the threshold blowing ratio required for producing VRs, ranging from 0.22--1.1~\cite{Zhong2005, Jabbal2008, Ho2022}. Furthermore, the production and persistence of VRs is not limited to laminar crossflows. Ho et al.~\cite{Ho2022} simulated the effects of a circular synthetic jet in a turbulent crossflow and VRs were still observed, supporting the notion that VRs in the present experiment could remain coherent after passing through the turbulent boundary layer.

The trajectory of the VRs is seen to be convergent toward the center span due to the contraction in the flow (Fig.~\ref{fig:smoke_overhead}). The contraction of the bulk flow is attributed to the spanwise pressure gradient, which forces the outer, slower fluid inwards toward the faster fluid at the midspan~\cite{Machado2024}. The flow contraction is most apparent in the streaklines passing over the outer edges of the SJA array, however an inward trajectory is also observed in the VRs produced by the jets nearest to the midspan. This effect is highlighted in Fig.~\ref{fig:Qcrit}, as the VRs first appear at the $z/c=0.03$ plane, while at the midspan they are only observed downstream of $x/c\approx0.5$. This is due to the $z/c=0.03$ plane being closer to the SJA, thus capturing the VR shortly after formation, whereas the midspan plane only captures the VRs farther downstream as they begin to drift into the frame due to the flow contraction.

The $z/c=0.03$ plane in Fig.~\ref{fig:Qcrit} illustrates that the top of the VR expands downstream, while the bottom portion shrinks. This phenomenon is attributed to the lower portion of the VR, rotating counter-clockwise, being counteracted by the clockwise vorticity of the boundary layer, thus leading to its contraction. Conversely, the vorticity of the boundary layer promotes the expansion of the upper portion of the VR, which has a clockwise rotational sense. This effect aligns with findings from previous studies \cite{Zhong2005, Jabbal2008, Sahni2011, Vasile2013}, all of which attribute the deformation of VRs to the resident vorticity in the boundary layer.

Lastly, the bottom portion of the VR is seen to lag behind the top, forming a tilted VR. Experiments \cite{Zhong2005, Jabbal2008} and simulations \cite{Cheng2009, Ho2022} have been conducted to investigate the tilting tendency of VRs in a crossflow, their mixing properties, and their strong potential in flow control applications. VRs tend to stretch and tilt when subjected to a shear flow, and the longer a VR is exposed to a shear layer, the more tilt is observed \cite{Jabbal2008}. Additionally, simulations showed that higher shear rates resulted in greater tilt angles~\cite{Cheng2009}. As VR tilting was also observed to occur outside the boundary layer, models~\cite{Zhong2005} have suggested the top and bottom portions of the VR are subject to Magnus forces in opposite directions due to their opposite rotation. Jabbal and Zhong~\cite{Jabbal2008} studied the effect of tilted VRs, and determined that they lead to heightened shear stresses, and thus enhanced fluid mixing. However, the enhanced mixing is attributed to tertiary vortex pairs near the wall, whose presence remains unverified in the present study due to turbulence and limitations in spatial resolution.

\begin{figure}
    \centering
    \includegraphics[width=0.5\linewidth]{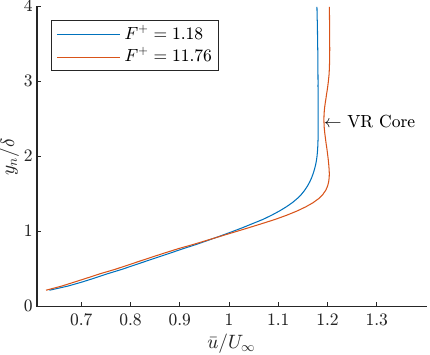}
    \caption{Mean velocity profiles at $x/c=0.6$, midspan}
    \label{fig:velocity_profiles}
\end{figure}

As a VR convects downstream, it decelerates fluid in its core while accelerating fluid around its periphery. Notably, the lower portion of the vortex ring possesses opposite vorticity as the boundary layer; thus, it tends to decelerate fluid above it while accelerating fluid beneath it, providing a mechanism for downward momentum transfer. Simultaneously, the opposite vorticity of the boundary layer tends to dissipate the vortex ring. Aligned with this analysis, a recent experimental study of circular SJAs~\cite{Xu2024} shows a strong correlation between vortex ring breakdown and entrainment enhancement. The influence of the VRs on the velocity field is most clearly illustrated by the mean velocity profiles, which are plotted in Fig.~\ref{fig:velocity_profiles} allowing for a comprehensive comparison between the two control cases. The VR core location is annotated on the plot, identified at $y/c=0.16$ in the global coordinate system. Low-frequency actuation at $F^+=1.18$ does not produce vortex rings, thus a typical velocity profile is observed. Conversely, a unique curve is observed for the $F^+=11.76$ case. A local minima in velocity coincides with the VR core, due to fluid in the VR core being ejected backward as it convects downstream. Below the velocity deficit exists an inflection point where the maximum velocity is achieved at $y_n/\delta=1.7$. This observation suggests that the VR is working as a mechanism for downward momentum transfer. The significance of this phenomena for effective flow control is further investigated with modal analysis in Sec.~\ref{sec:spanwise}.

\subsection{Off Center Span}
\label{sec:spanwise}

\begin{figure}
    \centering
    \includegraphics[width=0.5\linewidth]{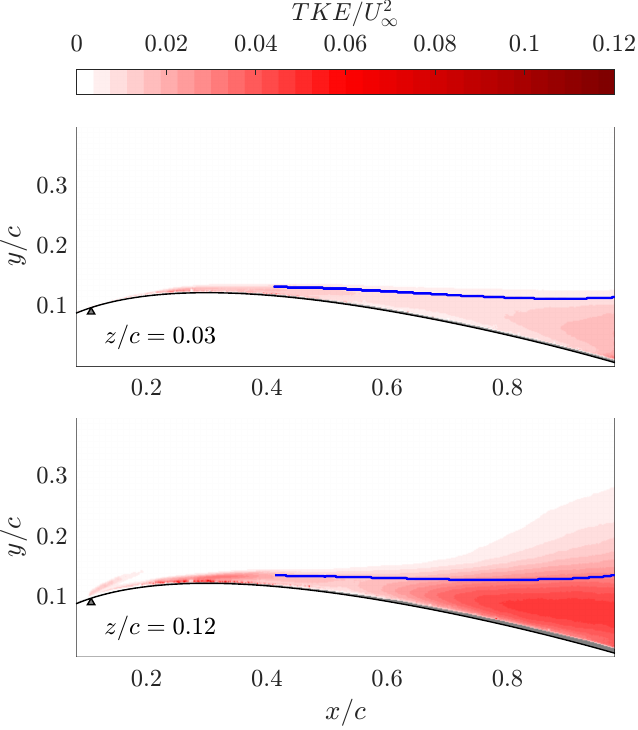}
    \caption{Contours of turbulent kinetic energy, $F^+=11.76$}
    \label{fig:TKE}
\end{figure}

Contours of turbulent kinetic energy (TKE) are plotted in Fig.~\ref{fig:TKE} for two spanwise planes: $z/c=0.03$ and $z/c=0.12$. Additionally, the solid blue line outlines the mean boundary layer for $x/c>0.4$. Near the midspan, at $z/c=0.03$, a turbulent boundary layer is observed, which grows downstream. The freestream flow over the airfoil, however, remains laminar. Farther from the midspan, at $z/c=0.12$, the turbulence near the wall intensifies, and is seen to propagate beyond the mean boundary layer. This indicates that the shear layer is highly unsteady at this spanwise location. The turbulent region grows especially rapidly downstream of $x/c=0.7$, indicating highly unsteady flow phenomena above the trailing edge of the airfoil. The heightened turbulence above the wing is indicative of unsteady aerodynamic forces, likely resulting in a lower sectional lift coefficient. Additionally, the upward extension of the turbulent region dissipates the vortex rings, explaining their absence at this spanwise location in Fig.~\ref{fig:Qcrit}.

\begin{figure}
    \centering
    \includegraphics[width=0.5\linewidth]{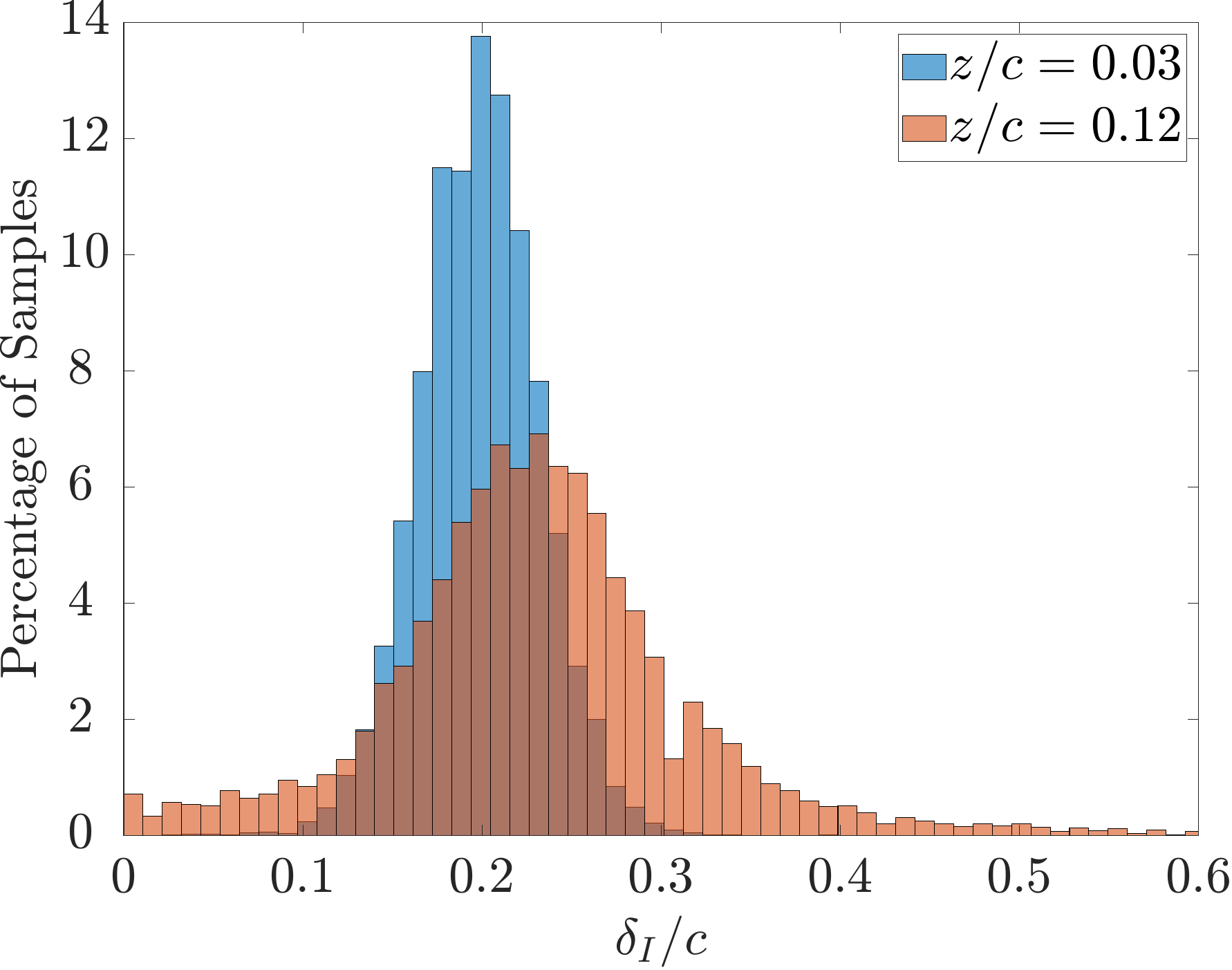}
    \caption{Distributions of instantaneous boundary layer thickness at $x/c=0.8$, $F^+=11.76$}
    \label{fig:BL_distributions}
\end{figure}

\begin{figure}
    \centering
    \includegraphics[width=0.5\linewidth]{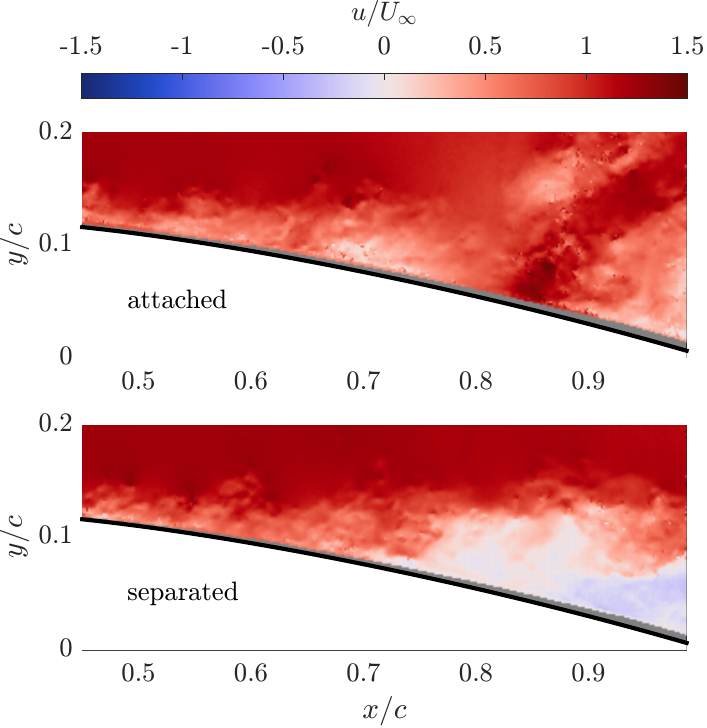}
    \caption{Contours of streamwise velocity for two selected instantaneous frames at $z/c=0.12$, $F^+=11.76$}
    \label{fig:attached_separated}
\end{figure}

Previous work~\cite{Machado2024} showed that differences in the boundary layer were not discernible from the smoke-wire visualizations up to $z/c=0.17$, and the present PIV results yielded only marginal variations in the mean flow. This emphasizes that the diminishing control authority away from the symmetry plane manifests initially in the flow's unsteadiness rather than its time-averaged characteristics. To further investigate the steadiness of the flow at different spanwise locations, distributions of the instantaneous boundary layer thickness, over all 8 phase angles, at $x/c=0.8$ are presented in Fig.~\ref{fig:BL_distributions}. The boundary layer distribution at $z/c=0.12$ is notably wider, indicating unsteady shear layer behavior. Previous experimental studies of mildly controlled flow over an airfoil with weak SJAs~\cite{Tang2014, Salunkhe2016} observed a ``flapping" shear layer which alternated between attached and separated states. Following their analysis, contours of the instantaneous streamwise velocity are plotted in Fig.~\ref{fig:attached_separated} for the plane $z/c=0.12$, illustrating both instances of an attached flow and a separated flow characterized by a reverse flow region near the trailing edge. At the $z/c=0.03$ plane, the data exhibited no instances of reverse flow, indicating that the control effects remain steady near the midspan.

\subsubsection{Modal Analysis}
\label{sec:modal}
The proper orthogonal decomposition (POD) is used to identify dominant unsteady features in the velocity fields at two spanwise planes. The resulting POD spatial modes highlight areas where velocity fluctuations are correlated, facilitating the identification of energetic coherent structures~\cite{Taira2017}. By analyzing these POD spatial modes, we aim to uncover the physical flow mechanisms that govern flow control, and how these mechanisms break down with increasing distance from the midspan. Phase locked PIV velocity data, sampled at $\phi=\SI{0}{\degree}$, is used in this modal analysis to ensure the high-frequency structures produced by the SJAs remain visible. The POD was repeated with the time-averaged velocity field, which yielded similar results with the exception of the SJA vortices appearing smeared, since their location is phase-dependent.

A fluctuating velocity field, $\mathbf{u'}(\mathbf{x},t)$ can be decomposed as

\begin{equation}
  \mathbf{u'}(\mathbf{x},t)=\sum_{k=1}^{\infty}a_k(t)\mathbf{\Phi}_k(\mathbf{x})
\end{equation}

where $\mathbf{\Phi}_k(\mathbf{x})$ are the spatial modes, and $a_k(t)$ are their time coefficients \cite{Chatterjee2000,Taira2017,Weiss2019}.

\begin{figure}
    \centering
    \includegraphics[width=0.5\linewidth]{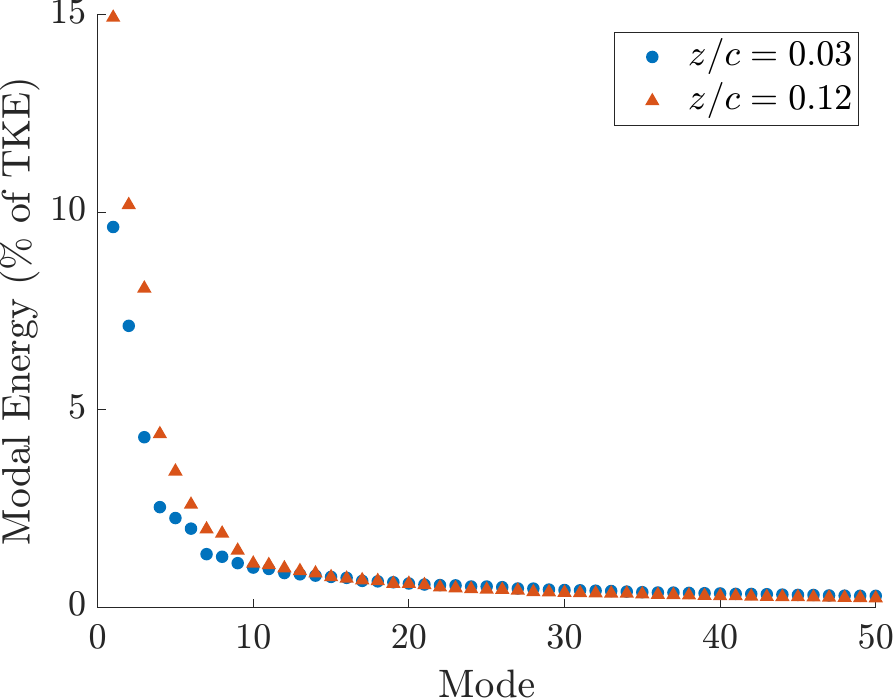}
    \caption{Fractional energy distributions for the first 50 modes, $F^+=11.76$}
    \label{fig:POD_energy}
\end{figure}

To perform POD on our discrete velocity field, the fluctuating velocity field $\mathbf{u'}(x,y,t)$ is flattened into a matrix, $A$, of dimensions $s \times t$, where $s$ is the number of spatial measurement locations sampled simultaneously, and $t$ is the number of time samples~\cite{Chatterjee2000}. A singular value decomposition is then performed on matrix $\mathbf{A}$ such that:

\begin{equation}
  \mathbf{A}=\mathbf{U\Sigma V^T}
\end{equation}

where:
\begin{samepage}
\begin{itemize}
    \item $\mathbf{U}$ is an orthogonal matrix of the spatial modes, arranged in descending order based on their contributions to the kinetic energy of the flowfield.
    \item $\mathbf{\Sigma}$ is a diagonal matrix containing the singular values of $\mathbf{A}$. Additionally the diagonal entities of $\mathbf{\Sigma}^2$ are proportional to the fractional turbulent kinetic energy associated with each spatial mode.
    \item $\mathbf{V^T}$ contains the random temporal coefficients of each spatial mode.
\end{itemize}
\end{samepage}
The fractional energy distribution of the first 50 POD modes in the high-frequency control case is plotted in Fig.~\ref{fig:POD_energy}. A comparison of the energy distributions at two spanwise planes shows that the first 14 modes at $z/c=0.12$ capture a higher proportion of the total fluctuating energy compared to those at $z/c=0.03$. This observation implies that, at spanwise locations farther from the midspan, larger structures contain a higher proportion of the total fluctuating energy. In contrast, near the midspan, the energy is more distributed across smaller scales, suggesting the effective dissipation of large-scale structures near the midspan. As the control authority diminishes farther from the symmetry plane, large-scale structures gain prominence in the flow.

\begin{figure*}
    \centering
    \includegraphics[width=\linewidth]{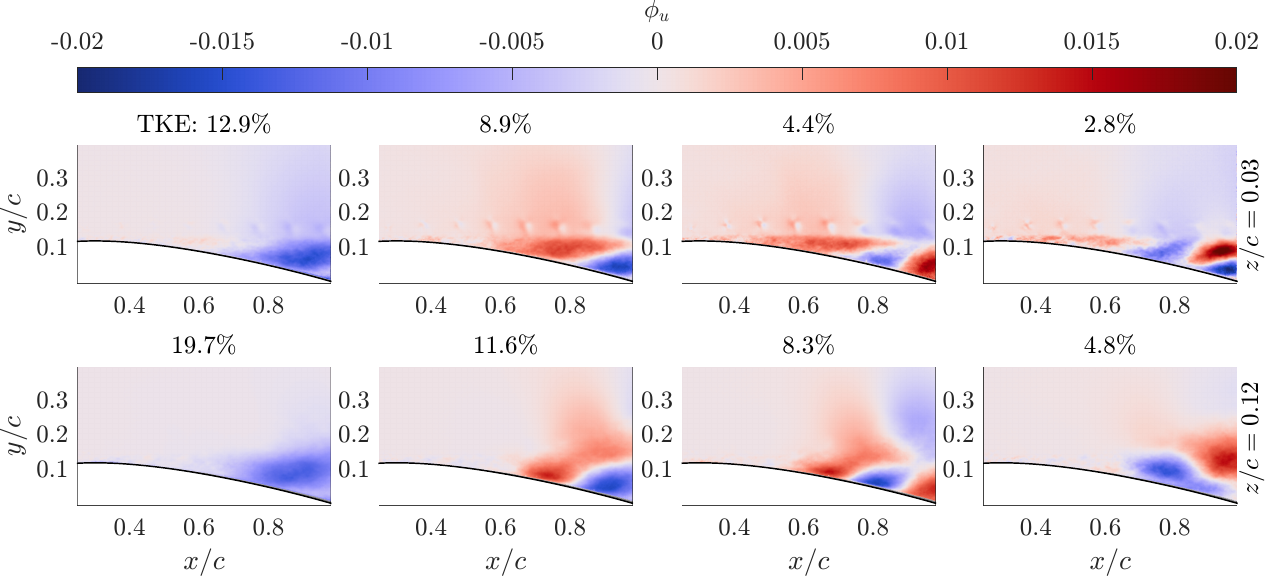}
    \caption{First 4 POD modes of the fluctuating streamwise velocity field for $z/c=0.03$ (top) and $z/c=0.12$ (bottom), $F^+=11.76$}
    \label{fig:uPOD}
\end{figure*}

\begin{figure*}
    \centering
    \includegraphics[width=\linewidth]{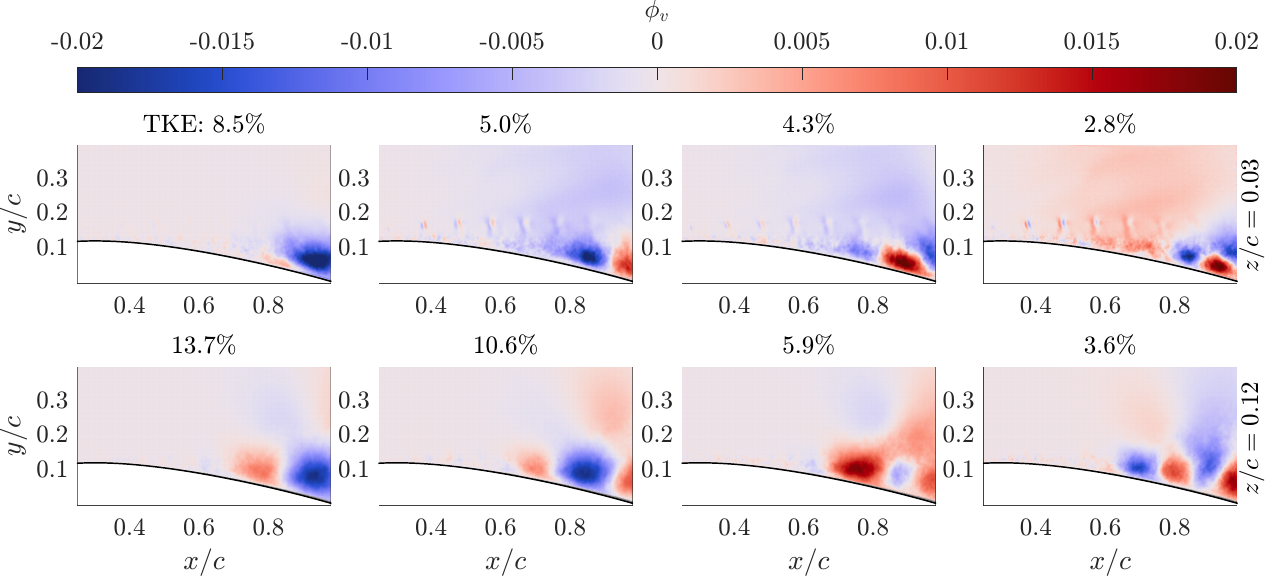}
    \caption{First 4 POD modes of the fluctuating transverse velocity field for $z/c=0.03$ (top) and $z/c=0.12$ (bottom), $F^+=11.76$}
    \label{fig:vPOD}
\end{figure*}

The first four spatial modes of the fluctuating streamwise velocity field are plotted in Fig.~\ref{fig:uPOD} for two spanwise locations, $z/c=0.03$ and $z/c=0.12$. Likewise, the spatial modes of the fluctuating transverse velocity field are plotted in Fig.~\ref{fig:vPOD}. In the first streamwise mode, a highly correlated area can be seen above the aft portion of the airfoil. This mode effectively captures the unsteady boundary layer region, which is most prominent downstream of $x/c\approx0.7$. The first transverse mode for both spanwise locations (Fig.~\ref{fig:vPOD}) reveals a concentrated, highly correlated area above the trailing edge region. This spatial mode suggests that variations in downwash have a synchronized local effect that is most energetic near the trailing edge. This observation aligns with the impact of the adverse pressure gradient, causing the fluid to decelerate as it moves downstream --- resulting in an area dominated by fluctuations. The fluctuations observed in the streamwise velocity are dependent on the amount of downwash near the trailing edge, a concept aligned with the phenomenon of downward momentum transfer from the free stream into the shear layer~\cite{Greenblatt2000,Amitay2002,Glezer2005,Salunkhe2016,Feero2017b,Kim2022,Yang2022,Xu2023}.

The notable difference in the first transverse mode between the spanwise locations is that the $z/c=0.03$ plot exhibits a single dominant mode, while the $z/c=0.12$ plot displays a pair of negatively correlated regions, indicative of a vortex. For example, a stronger clockwise vortex would result in a region of higher $v'$ followed by a region of greater $-v'$ downstream, hence the two negatively correlated areas. Spanwise vortices form over airfoils through the roll-up of the shear layer, stemming from the Kelvin-Helmholtz instability~\cite{Lang2004,Burgmann2008,Yarusevych2008,Boutilier2012,Kirk2017,Ziade2018,Machado2024}. The flow field at $z/c=0.12$ is characterized by a flapping shear layer, alternating between attached and separated states, while the flow at $z/c=0.03$ exhibited no instances of reverse flow. This indicates that while shear layer roll-up is suppressed at the midspan, it becomes more prevalent with increasing distance from the midspan, bearing resemblance to the baseline flow. In the current study, diminished control away from the midspan exhibits similar flow patterns to a mildly controlled flow with weak SJAs~\cite{Tang2014,Salunkhe2016}. Common features include the flapping tendency of the shear layer and the presence of shear layer roll-up.

The second streamwise mode (Fig.~\ref{fig:uPOD}) reveals two overlapping negatively correlated areas, downstream of $x/c\approx0.6$. This mode describes a large-scale vortex passing by this location. In the $z/c=0.03$ plot, this vortex appears elongated in the streamwise direction and remains swept to the airfoil surface. This vortex orientation is desirable as it results in a thinner shear layer and a narrower wake. The VRs identified previously in Section~\ref{VRs}, appear as four distinct red dots above the red structure (at $y/c\approx0.15$), suggesting that fluctuations in the near wall region are correlated with the strength of the VRs. A possible explanation is that the lower counterclockwise rotating part of the VR works to energize the near wall region by enhancing mixing, ultimately accelerating the fluid beneath it. This effect was demonstrated experimentally on a flat plate, which showed that SJA induced VRs produced fuller boundary layer profiles near the wall due to increased mixing with the freestream~\cite{Jabbal2008}. Though this mechanism is well proven in simple flow conditions, it is uncertain whether its effect is significant in the present complex flow over the airfoil. In the $z/c=0.12$ plot, the structure leaks higher in the $y$ direction, indicating that the fluctuations are not as controlled. This deteriorated control state could be due to the lack of the VR influence, however establishing causality remains challenging. From an aerodynamics perspective, the fluctuations higher above the airfoil surface impact the global flow field, resulting in a wider wake. The second transverse mode for the $z/c=0.03$ measurement plane (Fig.~\ref{fig:vPOD}) displays two adjacent negatively correlated patches starting at $x/c\approx0.8$. Similar to the first mode at $z/c=0.12$, this suggests the presence of large vortical structures as the fluctuating transverse velocities would be related inversely on the upstream and downstream sides of the vortex. However, it is noted that at the midspan, these vortices carry much less energy and are likely less frequent. VR interaction is also observed in this mode, suggesting that their presence aids in suppressing the trailing edge flow structures, which would help keep the wake narrow and uniform. In the $z/c=0.12$ plot, three alternating patches are visible, suggesting that large vortical structures develop further upstream. The formation of vortices farther upstream is attributed to a less favorable pressure gradient at this spanwise plane. A similar phenomenon was noted in an experimental study~\cite{Yarusevych2006} in which an increase in the angle of attack or a decrease in Reynolds number shifted the roll-up of the shear layer farther upstream.

The third and fourth modes show similar phenomena as seen in the second mode, but on a smaller scale. In each streamwise mode (Fig.~\ref{fig:uPOD}), the structures are seen to be elongated in the streamwise direction, and exist underneath the layer of VRs for the $z/c=0.03$ plane. Conversely, at $z/c=0.12$, the influence of the structures is observed to leak higher up into the freestream, negatively impacting the wake dynamics. Similarly, the transverse modes (Fig.~\ref{fig:vPOD}) display structures that appear closer to the surface at $z/c=0.03$, while the structures appear less controlled in the $z/c=0.12$ plane as they expand both upward and upstream. Every streamwise mode, and transverse modes \numrange{2}{4} of the $z/c=0.03$ plane display interactions between the VRs and the structures in the shear layer. This relationship provides supporting evidence for the role of VRs in downwash and momentum transport, influencing boundary layer behavior at various scales and energy levels.

\section{Conclusion}
In this study, an array of SJAs is used to reattach the flow over a stalled airfoil. The stability, flow structures, and three-dimensionality of the controlled flow are investigated experimentally for two actuation frequencies. Results show that while both low- and high-frequency actuation are effective in controlling flow separation and improving aerodynamic performance, high-frequency actuation at $F^+=11.76$ is particularly advantageous for creating a more stable flow. Low-frequency control at $F^+=1.18$ produced large-scale, periodic flow structures in the shear layer and wake, leading to unsteady aerodynamic forces. In contrast, high-frequency control effectively dissipated the large-scale flow features present in both the baseline and low-frequency control cases, resulting in a steady reattachment of the flow and superior aerodynamic performance.

Next, we identified flow structures induced by high-frequency actuation, and assessed their significance in flow control. Small scale structures were detected at the shear layer boundary, and were identified as VRs produced every actuation cycle. The dynamics of the VRs were analyzed providing insights into their evolution and trajectory as they convect over the airfoil. The rotational sense of the VR tends to promote momentum transfer from the freestream into the shear layer, which is a key mechanism in flow control. Additionally, modal analysis revealed a strong correlation between the VRs and structures in the shear layer, suggesting an interaction.

The spanwise control authority of the SJA array is evaluated by studying the flowfield both at and away from the midspan. The diminishing control efficacy away from the symmetry plane was first observed in the flow's unsteadiness rather than its time-averaged characteristics, the latter exhibiting minimal differences. The turbulent fluctuations in the shear layer intensify at spanwise planes further from the midspan, and expand above the mean boundary layer. Analysis of the instantaneous flowfield revealed the presence of a flapping shear layer, alternating between being attached and separated at only a modest distance from the midspan. This indicates a substantial reduction in the sectional lift coefficient away from the midspan.

Modal analysis is used to provide additional insights into the spanwise control authority and the significance of VRs in flow control. Near the midspan, vortical structures in the shear layer remained beneath the layer of VRs and were effectively kept swept to the wing. However, farther from the midspan, these structures formed further upstream and leaked higher up, resulting in a large and unsteady shear layer, with an absence of VRs. Lastly, while the control suppressed shear layer roll-up at the midspan, roll-up was detected for the plane away from the midspan, similar to the baseline case.

\begin{acknowledgments}
The authors acknowledge the support of the Natural Sciences and Engineering Research Council of Canada (NSERC) and the Digital Research Alliance of Canada.
\end{acknowledgments}

\section{AUTHOR DECLARATIONS}
\textbf{Conflict of Interest}

The authors have no conflicts to disclose.

\section*{Data Availability Statement}
The data that support the findings of this study are available from the corresponding author upon reasonable request.

\bibliographystyle{unsrt}
\bibliography{reference}

\end{document}